\documentclass[sn-basic]{sn-jnl}

\usepackage{graphicx}%
\graphicspath{{../figures/}{./}}%
\usepackage{amsmath,amssymb,amsfonts}%
\usepackage{amsthm}%
\usepackage[title]{appendix}%
\usepackage[ruled]{algorithm}%
\usepackage{algpseudocode}%
\newcommand{\scdots}{\cdot\!\cdot\!\cdot}

\begin{document}

\title[NMF-FFB]{NMF-FFB: Non-negative matrix factorization with feedforward-feedback structure}

\author[1]{\fnm{Kenichi} \sur{Satoh}\thanks{ORCID: \url{https://orcid.org/0000-0003-4436-9347}}}

\email{kenichi-satoh@biwako.shiga-u.ac.jp}

\affil[1]{\orgdiv{Faculty of Data Science}, \orgname{Shiga University}, \orgaddress{\street{Banba 1-1-1}, \city{Hikone}, \postcode{522-8522}, \state{Shiga}, \country{Japan}}}

\abstract{
Non-negative matrix factorization (NMF) approximates a non-negative
endogenous data matrix as $Y_1 \approx XB$, with non-negative latent
components $X$ and coefficients $B$. Standard covariate-aware NMF is
feedforward: $B$ depends only on exogenous variables $Y_2$, with no
latent feedback among endogenous variables. We propose NMF-FFB (NMF
with feedforward-feedback structure), an exploratory data-fitting
framework that embeds the simultaneous equation
$B = \Theta_1 Y_1 + \Theta_2 Y_2$ in NMF, where $\Theta_1$ is
non-negative latent feedback and $\Theta_2$ non-negative exogenous
pathways. NMF-FFB is positioned within data-fitting structural
equation modeling (SEM): it fits $Y_1$ directly rather than a
model-implied covariance, and is not a confirmatory measurement model
or a replacement for maximum-likelihood SEM under standard
confirmatory factor analysis assumptions. When $\rho(X\Theta_1)<1$,
the reduced form $Y_1 \approx (I-X\Theta_1)^{-1} X\Theta_2 Y_2$
defines a latent Leontief inverse separating direct from cumulative
feedback-amplified effects. Estimation uses regularized multiplicative
updates with orthogonality and sparsity penalties; an $X$-fixed
bootstrap summarizes uncertainty for the feedback spectral radius,
the amplification ratio, and path coefficients. Unlike conventional
SEM, NMF-FFB requires only the latent rank $Q$ and lets $X$ group
endogenous indicators into latent factors. This suits non-negative
additive data, automatic loading discovery, Leontief-type cumulative
effects, and small samples where covariance-based maximum-likelihood
fitting is ill-conditioned. Applications to Holzinger-Swineford,
Los Angeles pollution-mortality, and Mississippi county-level health
data demonstrate interpretable parts-based representations across
distinct latent-feedback regimes.
}


\maketitle

\section{Introduction}\label{sec1}

\subsection*{Non-negative matrix factorization and its extensions}

Non-negative matrix factorization (NMF) decomposes a high-dimensional
non-negative data matrix into a product of two non-negative factors,
producing additive, parts-based representations that are directly
interpretable \citep{lee1999,lee2000,cichocki2009,gillis2014}.
Driven by this interpretability, NMF has been extended along several
directions, including regression-style formulations that incorporate
exogenous covariates into the coefficient matrix
\citep{satoh2025var,satoh2025lab} and self-expressive
variants that encode relational structure among observations
\citep{galeano2022,galeano2024}.
These extensions, however, remain strictly feedforward in their
latent structure: an observation matrix
$Y_1\!\in\!\mathbb{R}_+^{P_1\times N}$ (rows = $P_1$ endogenous
variables, columns = $N$ samples) is explained by non-negative
components $X\!\in\!\mathbb{R}_+^{P_1\times Q}$ ($Q$ latent factors)
and a coefficient matrix $B\!\in\!\mathbb{R}_+^{Q\times N}$ driven by
exogenous variables $Y_2\!\in\!\mathbb{R}_+^{P_2\times N}$ alone,
with no mechanism for the endogenous variables to influence each
other through the latent components.
In many application domains---economic transactions, ecological
flows, epidemiological pathways, and health-outcome networks---the
endogenous variables are not merely outputs but also propagate
influence back through the system. Our starting point is therefore to
ask: can NMF be extended so that the coefficient matrix itself is
governed by a latent simultaneous-equation system?

\subsection*{Structural equation modeling via data-matrix fitting}

Simultaneous equations with latent variables are the traditional
domain of structural equation modeling (SEM). The classical
\emph{Covariance-Based SEM} approach fits a model-implied covariance
matrix to the sample covariance matrix \citep{zhang2016} and is
typically formulated for recursive (acyclic) path diagrams with
signed coefficients; we refer to it simply as ``SEM'' when the
context is clear. In parallel, a family of \emph{data-fitting}
alternatives has emerged that fits the observation matrix $Y_1$
directly: generalized structured component analysis
\citep[GSCA;][]{hwang2004}, PLS path modeling \citep{tenenhaus2005},
and, most recently, matrix decomposition SEM
\citep[MD-SEM;][]{yamashita2024}. These methods represent latent
quantities as linear combinations of observed variables and estimate
parameters by minimizing the reconstruction error of $Y_1$.

Our NMF-FFB can be understood alongside this data-fitting SEM family,
while extending it along three distinct axes.
First, \emph{non-negativity} is imposed on the components $X$ and the
structural coefficients $\Theta_1,\Theta_2$, which none of GSCA,
PLS-PM, or MD-SEM requires; this yields additive, parts-based path
interpretations and a Leontief-type cumulative reading rather than the
usual signed-coefficient interpretation of data-fitting SEM.
Second, \emph{cyclic latent feedback} is permitted: where GSCA,
PLS-PM, and MD-SEM assume recursive directed acyclic graph (DAG)
structures, NMF-FFB allows
$\Theta_1\ne 0$ and models loops among endogenous variables through
the non-negative latent components.
Third, the method is \emph{exploratory}: the latent path pattern is
learned from data rather than pre-specified, complementing the
confirmatory emphasis of the existing data-fitting methods and making
NMF-FFB applicable when no theoretical path diagram is available.

Although NMF is fundamentally a matrix-factorization method for
approximation, its non-negativity constraint gives the decomposition an
interpretable additive structure, so the fitted factors and paths can
be read as composition, propagation, and cumulative feedback patterns
in the data.

Classical nonrecursive SEM has long allowed reciprocal relations and
feedback loops under covariance-based, typically confirmatory
specifications \citep{bollen1989,paxton2011}.  NMF-FFB does not replace
that tradition; rather, it addresses a different use case by combining
non-negativity, exploratory data-matrix fitting, and latent feedback in
a single framework, thereby yielding parts-based components and a
Leontief-type cumulative interpretation that are not the usual target of
classical nonrecursive SEM.
Relatedly, non-negative structural systems have been studied in the
covariance- and DAG-based literature \citep{lodhia2023,zhou2022}, but
none of these jointly addresses the three features above.

To avoid ambiguity, the scope and non-goals of NMF-FFB should be stated
explicitly. It is not a covariance-based SEM in the classical sense: it
does not posit a model-implied covariance matrix such as
$\Sigma=\Lambda\Phi\Lambda^\top+\Theta_\epsilon$ and fit it to a sample
covariance matrix. It is also not a confirmatory measurement model, since
the loadings in $X$ are estimated from the data rather than specified
before fitting, and it is not intended to replace maximum-likelihood SEM
when standard CFA assumptions and a well-supported path diagram are
available. The intended target is narrower: exploratory non-negative data
settings in which latent loadings must be discovered, Leontief-type
cumulative effects are substantively meaningful, additive non-negative
components are part of the interpretation, or a conventional
covariance-based SEM benchmark is numerically fragile or requires strong
dataset-specific identifying restrictions.

\subsection*{From latent feedback to blind input--output analysis}

A notable consequence of introducing a non-negative cyclic structure
into data-fitting SEM is that the equilibrium of the resulting
simultaneous equation system has a familiar economic reading. When
the system $B = \Theta_1 Y_1 + \Theta_2 Y_2$ with $Y_1 \approx XB$ is
solved under the stability condition $\rho(X\Theta_1)<1$, the
reduced-form mapping
$Y_1 \approx (I-X\Theta_1)^{-1} X\Theta_2 Y_2$
admits an analogy with Leontief's input--output (IO)
equilibrium \citep{leontief1936}: a non-negative coefficient matrix
propagates exogenous demands into endogenous outputs, and the inverse
$(I-X\Theta_1)^{-1}$ plays the role of a latent analogue
of the Leontief inverse, whose Neumann expansion separates direct from
cumulative effects.

This IO interpretation was not the motivation for the model; it
emerged as a by-product of combining non-negativity with cyclic
feedback in the latent space.

Once this correspondence is recognized, NMF-FFB addresses what we
term the \emph{blind input--output problem}: recovering the
equilibrium propagation between exogenous and endogenous variables
when intermediate flows are unobserved and the internal coefficient
matrix is hidden. Classical IO requires full
observation of
transactions, whereas the blind variant---natural in health,
environmental, and cognitive applications---must infer the latent
propagation structure from aggregate observations alone.

\subsection*{Proposed method and key definitions}

Formally, NMF-FFB parameterizes the NMF coefficient matrix $B$ through
a non-negative simultaneous equation system
\[
B = \Theta_1 Y_1 + \Theta_2 Y_2,
\qquad X,\Theta_1,\Theta_2 \ge 0,
\]
coupled with the NMF approximation $Y_1 \approx XB$, so that the
endogenous observations $Y_1$ arise from non-negative latent
components $X$ driven by both exogenous inputs $Y_2$ and endogenous
feedback through $Y_1$ itself. Two central terms are defined as
follows.
\begin{itemize}
\item \textbf{Blind input--output analysis.}
      The problem of recovering equilibrium propagation from
      exogenous to endogenous variables when intermediate flows are
      unobserved.
\item \textbf{Latent feedback.}
      Cyclic dependence among endogenous variables mediated by
      latent non-negative components, quantified by the spectral
      radius $\rho(X\Theta_1)$.
\end{itemize}
As in MD-SEM \citep{yamashita2024}, the individual exploratory components
$(X,\Theta_1,\Theta_2)$ are not uniquely identifiable in general; the
\emph{equilibrium mapping}
$M_{\mathrm{model}} = (I-X\Theta_1)^{-1}X\Theta_2$
(formally introduced in Section~\ref{subsec:equilibrium})
is taken as the primary object of
inference, and its empirical stability is documented in
Sections~\ref{subsec:simulation}--\ref{subsec:comparison}
via the structural correlations $\mathrm{SC}_{\mathrm{map}}$ and
$\mathrm{SC}_{\mathrm{cov}}$ (defined in
\S\ref{subsec:SCmap}--\S\ref{subsec:SCcov}) and the bootstrap intervals of
Section~\ref{subsec:feedback_metrics}.

\medskip
\noindent\textit{The main contributions are as follows.}
\begin{enumerate}
\item[(1)] We formulate a non-negative structural system that admits
\emph{cyclic} latent feedback ($\Theta_1 \ne 0$)---a feature absent from
the recursive data-fitting SEM family---by embedding a simultaneous
equation system into the coefficient matrix of NMF.
\item[(2)] The column-stochastic basis $X$ doubles as a
\emph{data-driven measurement model}: correlated endogenous variables
are automatically grouped into latent factors that admit
construct-level interpretation, without requiring pre-specification
of factor loadings. Combined with the non-negative $L_1$ penalty on
$\Theta_2$, this design discourages sign-cancelling coefficient inflation
and yields interpretable exogenous pathways under multicollinearity.
\item[(3)] We derive an equilibrium input--output mapping
$M_{\mathrm{model}} = (I-X\Theta_1)^{-1}X\Theta_2$ whose Neumann
expansion separates direct from cumulative effects and is
summarized by an amplification ratio (AR; defined in
\S\ref{subsec:arbounds}).
\item[(4)] We formulate a blind input--output reading in the latent
space, supporting recovery of Leontief-type propagation
\citep{leontief1936} when intermediate flows are unobserved; the
Mississippi application in \S\ref{subsec:mshealth} concretely
illustrates this reading on county-level health-risk data.
\item[(5)] We develop a regularized multiplicative-update estimation
algorithm with orthogonality and sparsity penalties, evaluation
metrics tailored to equilibrium behavior rather than covariance fit,
and an $X$-fixed bootstrap procedure that delivers 95\% percentile
intervals for the feedback diagnostics $\rho(X\Theta_1)$ and AR and
one-sided support-rate significance markers for the individual
non-negative path coefficients in $\Theta_1$ and $\Theta_2$, providing
path-level support markers that can be read alongside the significance
markers reported by standard SEM packages.
\end{enumerate}

Section~\ref{sec2} develops the model and equilibrium representation.
Section~\ref{sec3} introduces the regularized multiplicative-update
algorithm for estimating $(X,\Theta_1,\Theta_2)$.
Section~\ref{sec4} presents equilibrium-based structural metrics.
Section~\ref{sec5} demonstrates NMF-FFB on cognitive, environmental,
and population-health applications. Section~\ref{sec6} concludes.

\medskip
\noindent\textit{Dataset abbreviations.} Throughout the simulation and
empirical sections we use the following short labels for the three
application datasets: \emph{HS39} (Holzinger--Swineford 1939 cognitive
battery; \S\ref{subsec:hs39}), \emph{LA} (Los Angeles
pollution--mortality; \S\ref{subsec:pollution}), and
\emph{Mississippi} (county-level health-risk data;
\S\ref{subsec:mshealth}). The simulation tables in \S\ref{sec4} use
the HS39 dimensions $(P_1,P_2,Q)=(9,3,3)$ as a reference design so
that the simulation diagnostics are directly comparable to the HS39
empirical results.

\section{The NMF-FFB Model}\label{sec2}

This section develops the NMF-FFB framework as a non-negative
feedforward-feedback factorization model.
The goal is to enrich the parts-based representation of NMF by imposing a
non-negative simultaneous equation system on the coefficient matrix.
This structural layer induces interpretable direct and feedback effects,
and yields an equilibrium mapping that admits an analogy with
classical input--output (IO) analysis.

\subsection{Basic formulation}\label{subsec:basic}

\paragraph*{Notation.}
Throughout the paper we use the following notation, collected here for
reference:
$N$ denotes the sample size (number of observational units, arranged in
columns);
$P_1$ the number of endogenous (outcome) indicators and
$P_2$ the number of exogenous (covariate) indicators;
$Q$ the number of non-negative latent factors.
$Y_1\in\mathbb{R}_+^{P_1\times N}$ and $Y_2\in\mathbb{R}_+^{P_2\times N}$
are the non-negative data matrices of endogenous and exogenous variables,
respectively; in the applications below they are rescaled to $[0,1]$, but
the formal model requires only non-negativity.
$X\in\mathbb{R}_+^{P_1\times Q}$ is the basis (loading) matrix whose
columns sum to one and represent latent profiles over the endogenous
variables;
$\Theta_1\in\mathbb{R}_+^{Q\times P_1}$ and
$\Theta_2\in\mathbb{R}_+^{Q\times P_2}$ are the structural-coefficient
matrices that activate latent factors from $Y_1$ (feedback) and $Y_2$
(exogenous drivers), respectively; and
$B = \Theta_1 Y_1+\Theta_2 Y_2\in\mathbb{R}_+^{Q\times N}$ is the
resulting matrix of latent mixture weights used in the NMF
factorization $Y_1\approx XB$.
We adopt the variables-$\times$-individuals convention (rows index
variables, columns index individuals) so that $X$ and the structural
coefficient matrices act naturally by left-multiplication on the
data, following the original NMF formulation
\citep{lee1999} and the document-term convention common in the
matrix-factorization literature.

Recall the NMF approximation $Y_1 \approx XB$ with column-stochastic
$X$. Classical NMF treats $B$ as unconstrained; NMF-FFB instead
specifies $B$ through a structural system.

\subsection{Structural equation component}\label{subsec:structural}

We model the coefficient matrix $B$ as a non-negative simultaneous
equation system:
\[
B = \Theta_1 Y_1 + \Theta_2 Y_2,
\qquad
\Theta_1 \in \mathbb{R}_+^{Q \times P_1},\;
\Theta_2 \in \mathbb{R}_+^{Q \times P_2}.
\]
Here, $\Theta_1$ governs how endogenous variables activate the latent components (feedback),
and $\Theta_2$ governs how exogenous drivers activate latent components.

Substituting this expression for $B$ into $Y_1 \approx XB$ gives the
structural form of NMF-FFB:
\begin{equation}
Y_1 \approx X(\Theta_1 Y_1 + \Theta_2 Y_2).
\label{eq:structural}
\end{equation}
The composite matrices $X\Theta_1$ and $X\Theta_2$ encode,
respectively, direct endogenous interactions and direct exogenous
influences in the observed space.

The special case $\Theta_1 = 0$ reduces to the feedforward
NMF-with-covariates family of \citet{satoh2025var,satoh2025lab}, in
which $Y_1 \approx X\Theta_2 Y_2$ has no latent feedback channel.
Allowing $\Theta_1 \neq 0$ introduces latent feedback and leads to an
equilibrium representation.

\paragraph*{Identifiability, score uniqueness, and absence of variable uniquenesses.}
Three aspects of NMF-FFB deserve explicit clarification in the SEM context.
\begin{itemize}
\item \textbf{Rotational indeterminacy.} Under non-negativity,
the pair $(X,B)$ is no longer subject to the full orthogonal
rotational indeterminacy of Gaussian factor analysis: arbitrary
rotations are excluded by the non-negativity constraint, and the
column-stochastic normalization of $X$ (each column sums to one,
cf.~\S\ref{subsec:basic}) further reduces the positive-rescaling
ambiguity, with label-permutation among the $Q$ factors as the
remaining symmetry at the level of these continuous indeterminacies.
This does not, however, imply uniqueness of the NMF solution itself:
\emph{global uniqueness} generally requires additional structural
conditions on $(X,B)$ such as separability or sufficient spread of
the columns of $X$ \citep{Donoho2003,Laurberg2008}.
We therefore claim only that NMF-FFB inherits the partial
identifiability of NMF in this conditional sense---suppressing the
principal continuous indeterminacies but not guaranteeing global
uniqueness in general.
\item \textbf{Score uniqueness.} Given estimates of $X,\Theta_1,\Theta_2$,
the latent score matrix $B=\Theta_1 Y_1+\Theta_2 Y_2$ is a determined
(closed-form) function of the observed $Y_1,Y_2$. There is thus no
separate score-estimation step and none of the factor-score
indeterminacy of Covariance-Based SEM.
\item \textbf{Variable uniquenesses.} NMF-FFB is a data-matrix-fitting
model (in the tradition of GSCA, PLS-PM, and MD-SEM); there is no
residual-variance ``uniqueness'' parameter attached to each observed
variable. We state this as a deliberate design choice rather than an
oversight; the cost of fit is the reconstruction residual, discussed in
\S\ref{subsec:gof}.
\end{itemize}

\subsection{Equilibrium representation}\label{subsec:equilibrium}

Rearranging \eqref{eq:structural} yields
\[
(I - X\Theta_1)\, Y_1 \approx X\Theta_2 Y_2.
\]
To ensure that the system converges, we impose the stability condition
\begin{equation}
\rho(X\Theta_1) < 1,
\label{eq:stability}
\end{equation}
where $\rho(\cdot)$ denotes the spectral radius.
We say that the system is in \emph{equilibrium} when
\eqref{eq:stability} holds, so that
$(I-X\Theta_1)\,Y_1 \approx X\Theta_2\,Y_2$ admits a finite, unique
non-negative solution for $Y_1$; throughout this paper
``equilibrium representation'' and ``equilibrium system'' are used in
this sense.
Under this condition, $(I - X\Theta_1)$ is invertible, and the structural
system admits the reduced-form equilibrium mapping
\begin{equation}
Y_1 \approx (I - X\Theta_1)^{-1} X\Theta_2\, Y_2.
\label{eq:equilibrium}
\end{equation}

We refer to this matrix as the \emph{equilibrium mapping}:
\begin{equation}
M_{\mathrm{model}}
= (I - X\Theta_1)^{-1} X\Theta_2
\in \mathbb{R}_+^{P_1 \times P_2}.
\label{eq:Mmodel}
\end{equation}
The equilibrium mapping combines both direct effects and all rounds
of latent feedback. We use the term \emph{Leontief inverse} for the
factor $(I - X\Theta_1)^{-1}$ alone, reserving \emph{equilibrium
mapping} for $M_{\mathrm{model}}$.
In estimation, we enforce \eqref{eq:stability} and select hyperparameters only among stable solutions.

\subsection{Latent Leontief inverse and cumulative feedback}
\label{subsec:leontief}

Because $X\Theta_1$ is non-negative and satisfies \eqref{eq:stability},
the inverse in \eqref{eq:equilibrium} admits the Neumann expansion
\begin{equation}
(I - X\Theta_1)^{-1}
= I + (X\Theta_1) + (X\Theta_1)^2 + \scdots .
\label{eq:neumann}
\end{equation}
This is the latent analogue of the Leontief inverse in IO analysis.
Substituting \eqref{eq:neumann} into \eqref{eq:Mmodel} shows that
\[
M_{\mathrm{model}}
=
X\Theta_2
+ X\Theta_1 X\Theta_2
+ (X\Theta_1)^2 X\Theta_2
+ \scdots ,
\]
where each term corresponds to additional rounds of latent propagation.
All effects accumulate monotonically due to non-negativity.

A useful scalar summary of total amplification due to latent feedback is the amplification ratio:
\[
\mathrm{AR}
=
\frac{\|M_{\mathrm{model}}\|_{1,\mathrm{op}}}
     {\|X\Theta_2\|_{1,\mathrm{op}}} ,
\label{eq:amplification}
\]
where $\|\cdot\|_{1,\mathrm{op}}$ is the operator 1-norm.
Values above one indicate amplification generated by latent feedback.

\subsection{Amplification bounds}\label{subsec:arbounds}

Because $X\Theta_1$ is non-negative, the Leontief inverse
$(I - X\Theta_1)^{-1}$ admits the Neumann expansion of non-negative
powers of $X\Theta_1$ whenever $\rho(X\Theta_1)<1$. Consequently,
\[
M_{\mathrm{model}}
= (I - X\Theta_1)^{-1} X\Theta_2
\ge X\Theta_2 ,
\qquad 
\mathrm{AR} \ge 1,
\]
with equality in particular when $X\Theta_1=0$.

If $\|X\Theta_1\|_{1,\mathrm{op}}<1$, submultiplicativity yields
\[
\|(I-X\Theta_1)^{-1}\|_{1,\mathrm{op}}
\le \frac{1}{1-\|X\Theta_1\|_{1,\mathrm{op}}},
\]
and therefore the amplification ratio satisfies the two-sided bound
\[
1 \le \mathrm{AR}
\le \frac{1}{1-\|X\Theta_1\|_{1,\mathrm{op}}}.
\]
Thus AR remains finite under the stability condition and is non-decreasing
under elementwise increases in the feedback matrix $X\Theta_1$.
In particular, $\mathrm{AR}=1$ when $X\Theta_1=0$.
More generally, $\mathrm{AR}>1$ whenever the higher-order feedback terms
increase the maximal column sum of $X\Theta_2$.

\subsection{Interpretation of latent structure}\label{subsec:interpret}

The matrix $X$ provides a parts-based representation of $Y_1$, grouping
variables with similar response profiles into latent components.
The matrices $\Theta_1$ and $\Theta_2$ determine how these components are
activated:  
$\Theta_2$ allocates exogenous drivers to latent factors, and
$\Theta_1$ induces propagation through latent feedback.
The equilibrium mapping $M_{\mathrm{model}}$ summarizes how shocks to
$Y_2$ translate into equilibrium outcomes in $Y_1$.

Non-negativity ensures additive interpretation: increases in $Y_2$ can
only increase (or leave unchanged) the equilibrium values in $Y_1$, and feedback effects accumulate rather than offset.

\section{Estimation}\label{sec3}

This section outlines a practical estimation procedure for NMF-FFB.
Because the coefficient matrix is constrained by
\[
B=\Theta_1 Y_1 + \Theta_2 Y_2,
\]
the parameters $(X,\Theta_1,\Theta_2)$ must be estimated jointly under
non-negativity and the equilibrium structure introduced in
Section~\ref{sec2}.  
Our approach consists of:  
(i) initialization via a feedforward model,  
(ii) structural regularization,  
(iii) multiplicative updates, and  
(iv) hyperparameter selection based on equilibrium prediction error.

\subsection{Initialization}\label{subsec:init}

We adopt a two-phase warm start.
Phase 1 fits the feedforward special case $\Theta_1=0$, in which the
model reduces to the NMF-with-covariates tri-factorization
$Y_1 \approx X_0 \Theta_0 Y_2$ \citep{satoh2025var,satoh2025lab};
$X_0$ is initialized by NNDSVDar \citep{boutsidis2008}, a
deterministic SVD-based scheme that seeds $(X,B)$ from the leading
$Q$ singular triplets of $Y_1$ and replaces the remaining zero
entries by small positive values drawn from
$\mathrm{Uniform}(0,\bar{y}/100)$ (where $\bar{y}$ denotes the mean
of $Y_1$), so that entries are not frozen at the absorbing state
under multiplicative updates.
Phase 2 estimates the full $(X,\Theta_1,\Theta_2)$ system using $X_0$
as the $X$ initialization while $\Theta_1,\Theta_2$ are seeded with
small positive random entries (away from the absorbing boundary
$\Theta_k=0$); this warm start is intended to reduce, but cannot
eliminate, local-minimum failures that trap the fit at the stability
boundary $\rho(X\Theta_1)\to 1$.
The associated direct mapping $M_{\mathrm{simple}} = X_0\Theta_0$ is
later used as a structural benchmark when assessing IO fidelity in
Section~\ref{sec4}.

\subsection{Structural regularization}\label{subsec:regularization}

To reduce the remaining factor-overlap ambiguity and stabilize latent
feedback, we employ a penalized objective:
\[
L =
\|Y_1 - X(\Theta_1 Y_1 + \Theta_2 Y_2)\|_F^2
+ \frac{\lambda_X}{2}\|X^\top X - \mathrm{diag}(X^\top X)\|_F^2
+ \lambda_1\|\Theta_1\|_1
+ \lambda_2\|\Theta_2\|_1.
\]
The orthogonality penalty encourages separated latent components, while
the $\ell_1$ penalties promote sparse structural pathways and regularize
the feedback matrix $X\Theta_1$. Sparsity is introduced not only for
regularization but also for interpretation: by shrinking unnecessary
pathways toward the non-negative boundary, it reduces diffuse
cross-loadings and makes the latent components and feedback pathways
easier to read.

\paragraph*{$N$-scaled penalty convention.}
To guard against sample-size dependence in \emph{cross-$N$ or
cross-dataset} comparisons of the equilibrium diagnostics
$\rho(X\Theta_1)$ and $\mathrm{AR}$, we adopt the $N$-scaled convention
$(\lambda_X,\lambda_1,\lambda_2) \mapsto
(N/N_0)\,(\lambda_X,\lambda_1,\lambda_2)$
with reference size $N_0=50$, so that the effective regularization
strength remains comparable as $N$ grows. This convention is reserved
for cross-$N$/cross-dataset comparisons. Table~\ref{tab:sim00}
deliberately retains the original fixed-penalty protocol as a stress
test that exposes the sample-size effect; the $N$-scaled counterpart is
then reported separately in Section~\ref{subsubsec:Nscale}. For
single-dataset analyses (the real-data analyses in \S\ref{sec5}) we retain the
\emph{fixed-penalty} reading, which remains interpretable as an
ordinal descriptor of feedback strength within a fixed regularization
regime. This is a design choice, not a post-hoc correction; its
empirical validation is deferred to
Section~\ref{subsubsec:Nscale}.

\paragraph*{Rationale for the asymmetric choice of penalties.}
The orthogonality penalty is applied to $X$, whereas $\ell_1$ penalties
are applied to $\Theta_1$ and $\Theta_2$. This asymmetry is by design
and reflects what each matrix represents.
\begin{itemize}
\item $X$ is a \emph{basis matrix} whose columns represent distinct
latent factors; near-orthogonality of these columns is what makes the
bases interpretable as \emph{separate} parts of the data (the
``parts-based'' rationale of NMF; \citealp{lee1999}). An $\ell_1$
penalty on $X$ would drive individual entries toward zero but would
not enforce separation \emph{between} columns.
\item $\Theta_1$ and $\Theta_2$ are \emph{path-coefficient matrices},
for which element-wise sparsity is substantively meaningful---an
element shrunken to the non-negative boundary corresponds to an absent
path in a path diagram. The $\ell_1$ norm is the standard choice for
this purpose.
\item Imposing orthogonality on $\Theta_1$ or $\Theta_2$ would
constrain path coefficients to be mutually orthogonal as vectors,
which has no substantive interpretation in a path-diagrammatic model.
\end{itemize}
Thus the two penalty types play distinct roles (column separation for
$X$, element sparsity for $\Theta$) and are not interchangeable.

\subsection{Multiplicative updates}\label{subsec:updates}

Let $B=\Theta_1 Y_1 + \Theta_2 Y_2$ and $\hat{Y}_1 = X B$.  
Ignoring penalties, standard auxiliary-function arguments yield the
following multiplicative updates, which generalize the classical
Lee--Seung rules for NMF \citep{lee1999, lee2000}:
\[
X \leftarrow X \odot \frac{Y_1 B^\top}{XBB^\top},\quad
\Theta_1 \leftarrow \Theta_1 \odot \frac{X^\top Y_1 Y_1^\top}{X^\top \hat{Y}_1 Y_1^\top},\quad
\Theta_2 \leftarrow \Theta_2 \odot \frac{X^\top Y_1 Y_2^\top}{X^\top \hat{Y}_1 Y_2^\top}.
\]

Under regularization, the denominators incorporate the corresponding
gradients:
\[
X \leftarrow X\odot
\frac{Y_1 B^\top}
     {XBB^\top + \lambda_X X(X^\top X - \mathrm{diag}(X^\top X))},
\]
\[
\Theta_1 \leftarrow \Theta_1 \odot
\frac{X^\top Y_1 Y_1^\top}{X^\top \hat{Y}_1 Y_1^\top + \lambda_1},
\qquad
\Theta_2 \leftarrow \Theta_2 \odot
\frac{X^\top Y_1 Y_2^\top}{X^\top \hat{Y}_1 Y_2^\top + \lambda_2}.
\]

Columns of $X$ are renormalized after each update.
Appendix~\ref{app:derivation} gives the auxiliary-function derivation
for the reconstruction part and the KKT-type positive-gradient split used
for the regularized updates.
The post-update column renormalization is an implementation step that
fixes scale ambiguity of the factorization; convergence is therefore
monitored empirically by the stopping rule stated below rather than by
claiming monotone decrease of the fully regularized and renormalized
sweep.

\begin{algorithm}[htbp]
\caption{Regularized multiplicative updates for NMF-FFB.}
\label{alg:nmfffb}
\begin{algorithmic}[1]
\Require $(Y_1, Y_2)$, rank $Q$, penalties $(\lambda_X,\lambda_1,\lambda_2)$,
         tolerance $\varepsilon$ (set to $10^{-6}$ throughout this paper),
         max iterations $t_{\max}$ (set to $5000$ throughout this paper)
\State Initialize $X$ by NNDSVDar from the feedforward NMF-with-covariates
       fit \citep{boutsidis2008}
\State Initialize $\Theta_1,\Theta_2$ as small positive random matrices
\State $(X^\star,\Theta_1^\star,\Theta_2^\star)\gets(X,\Theta_1,\Theta_2)$;\quad
       $L^\star\gets+\infty$
\For{$t=1,2,\ldots,t_{\max}$}
  \State Update $X$, $\Theta_1$, $\Theta_2$ by the multiplicative rules
         (Appendix~\ref{app:derivation})
  \State Renormalize columns of $X$ to sum to one
  \State $L^{(t)}\gets$ penalized objective at $(X,\Theta_1,\Theta_2)$
  \If{$L^{(t)}<L^\star$}
    \State $(X^\star,\Theta_1^\star,\Theta_2^\star)\gets(X,\Theta_1,\Theta_2)$;\
           $L^\star\gets L^{(t)}$
  \EndIf
  \If{$t\ge 2$ \textbf{and} $|L^{(t)}-L^{(t-1)}|/\max(|L^{(t)}|,1)\le\varepsilon$}
    \State \textbf{break}
  \EndIf
\EndFor
\State \textbf{If} the loop exits via $t_{\max}$ rather than the
       convergence test, flag the replicate as non-convergent
\State Accept the fit only if $\rho(X^\star\Theta_1^\star)<1$;
       otherwise flag as non-stable
\State \Return $(X^\star,\Theta_1^\star,\Theta_2^\star)$
\end{algorithmic}
\end{algorithm}

The hyperparameters $(\lambda_X,\lambda_1,\lambda_2)$ are selected by
cross-validating the equilibrium mean absolute error (MAE), as defined
in \S\ref{subsec:crossvalidation}. Implementation details---the
explicit form of the stopping rule, the specific values of
$\varepsilon$ and $t_{\max}$ used in the numerical experiments, the
KKT-type stationarity argument, and caveats on non-convexity---are
deferred to Appendix~\ref{app:conv}.

\subsection{Hyperparameter selection}\label{subsec:crossvalidation}

Regularization parameters $(\lambda_X,\lambda_1,\lambda_2)$ control
orthogonality, sparsity, and feedback magnitude.  
Because reconstruction error does not reflect the equilibrium structure
introduced in Section~\ref{sec2}, hyperparameters are selected by 
cross-validating the equilibrium prediction performance using the
mapping $Y_1 \approx M_{\mathrm{model}} Y_2$ defined in
\eqref{eq:Mmodel}.  

For each held-out fold, equilibrium predictions are computed from the
estimated $M_{\mathrm{model}}$, and prediction accuracy is quantified by
the mean absolute error (MAE), defined as the average absolute deviation
across all entries of $Y_1$ and $\widehat{Y}_1$.
Specifically, the $N$ columns of $(Y_1,Y_2)$ are partitioned into $K$
disjoint folds; for each fold $k=1,\ldots,K$, the parameters
$(X^{(-k)},\Theta_1^{(-k)},\Theta_2^{(-k)})$ are fit by
Algorithm~\ref{alg:nmfffb} using only the training columns (those of
$(Y_1,Y_2)$ outside fold $k$), where the superscript $(-k)$ denotes a
quantity estimated on all folds excluding fold $k$, and the
training-fold operator
$M^{(-k)}_{\mathrm{model}}
= (I - X^{(-k)}\Theta_1^{(-k)})^{-1} X^{(-k)}\Theta_2^{(-k)}$
is then applied to the held-out $Y_2^{(k)}$ to produce predictions
$\widehat{Y}_1^{(k)} = M^{(-k)}_{\mathrm{model}} Y_2^{(k)}$.
The MAE is accumulated over all held-out entries across the $K$ folds.
We acknowledge that MAE serves both as tuning criterion and as a
headline reconstruction metric; cross-validation guards against in-sample
overfitting by evaluating prediction on columns unseen during estimation,
and Sections~\ref{subsec:simulation}--\ref{subsec:comparison}
complement MAE with the structural correlations
$\mathrm{SC}_{\mathrm{map}}$ and $\mathrm{SC}_{\mathrm{cov}}$ that are
not part of the tuning objective.  
In the empirical applications, the latent dimension $Q$ is selected
before this penalty grid by the same held-out equilibrium-MAE criterion
over feasible low-rank candidates; the selected $Q$ values are reported
in Table~\ref{tab:datasets}. Throughout this paper, $(Y_1, Y_2)$ are
min-max rescaled to $[0,1]$ (\S\ref{subsec:SCcov}); under this
preprocessing we use $\lambda_X=100$ as a working default, which
drives the off-ratio metric of \S\ref{subsec:ortho_check} below
$\approx 10^{-2}$ on all datasets considered in
\S\ref{sec4}--\S\ref{sec5}. The value of $\lambda_X$ that
achieves this off-ratio target depends on the data scale and on the
sample size $N$: when variables are not in $[0,1]$ or when $N$
deviates substantially from the working range of \S\ref{sec5},
$\lambda_X$ should be re-tuned to satisfy the off-ratio criterion
(the $N$-scaled convention introduced above provides one such
adjustment for cross-$N$ comparisons). With $\lambda_X$ fixed in
this way, $(\lambda_1,\lambda_2)$ are chosen by $K$-fold
cross-validation, subject to the stability requirement
$\rho(X\Theta_1)<1$.

\subsection{Relation to classical goodness-of-fit indices}\label{subsec:gof}

Standard SEM fit indices---GFI, CFI, TLI, RMSEA, AIC, BIC---are
built from the $\chi^2$ discrepancy between the sample and
model-implied covariance matrices, whose asymptotic distribution is
derived under maximum-likelihood estimation of a covariance
structure \citep{zhang2016}. NMF-FFB does not fit such a covariance
structure: following the data-fitting SEM tradition
\citep{hwang2004,tenenhaus2005,yamashita2024}, the penalized
objective $L$ measures the reconstruction of the observation matrix
$Y_1$ itself, so the associated discrepancy function has no
canonical $\chi^2$ limit. CFI, TLI, and RMSEA therefore have no
direct translation in NMF-FFB---just as they are ``not considered''
in PLS path modeling for the same reason---and AIC/BIC have no
consensus form here, since proposals differ in how to define the
effective degrees of freedom under non-negativity and
sparsity-inducing penalties \citep{schmidt2009,cai2023}.

This is not merely a definitional gap: classical covariance-based
indices, when computed naively from the implied second moment
$\Sigma_{\mathrm{model}}=M_{\mathrm{model}}\,S_{Y_2}\,
M_{\mathrm{model}}^{\top}+\mathrm{diag}(\widehat{\sigma}_\varepsilon^{2})$,
fail in three structural ways for the present applications.
First, the matrix-level parameter count of NMF-FFB
($P_1 Q + Q P_1 + Q P_2 + P_1$ for $X,\Theta_1,\Theta_2$ and
residual variances) exceeds the available unique covariance moments
$P_1(P_1{+}1)/2$ in all three applications of \S\ref{sec5},
producing negative degrees of freedom and pathological
$\chi^2$-derived indices.
Second, the fitted $X$ and $(\Theta_1,\Theta_2)$ optimize the
data-matrix Frobenius loss $\|Y_1-XB\|_F^{2}$ rather than the
covariance discrepancy, so even $df$-free indices such as the GSCA
GFI \citep{hwang2004} systematically underestimate NMF-FFB's
quality relative to a covariance-targeted ML estimator.
Third, the reduced-form prediction $M_{\mathrm{model}}Y_2$ contains
no per-variable intercept, by design of the non-negative
parts-based formulation; this absence further inflates residual
variance under any mean-aware comparison with classical SEM, which
explicitly estimates intercepts.
Accordingly, model quality is assessed by the suite of metrics
developed in Section~\ref{sec4}: the reconstruction error MAE, the
input--output structural correlations $\mathrm{SC}_{\mathrm{map}}$
and $\mathrm{SC}_{\mathrm{cov}}$, and the feedback diagnostics
$\rho(X\Theta_1)$ and $\mathrm{AR}$; hyperparameter selection uses
cross-validated equilibrium MAE
(Section~\ref{subsec:crossvalidation}).

\section{Evaluation}\label{sec4}

NMF-FFB is designed for structural interpretation.  
Accordingly, evaluation focuses on whether the estimated model
(1) preserves key input--output (IO) relationships,
(2) reproduces second-moment structure among endogenous variables,
(3) captures the strength of latent feedback, and
(4) yields a reasonable equilibrium approximation.
We summarize the corresponding metrics below.

\subsection{Input--output structural fidelity}\label{subsec:SCmap}

The equilibrium mapping
\[
M_{\mathrm{model}}=(I-X\Theta_1)^{-1}X\Theta_2
\]
summarizes the total effect of exogenous variables on endogenous ones.
To assess whether NMF-FFB preserves the qualitative IO structure implied
by the feedforward initialization, we compare $M_{\mathrm{model}}$ with
$M_{\mathrm{simple}} = X_0\Theta_0$ using the Pearson correlation
between their vectorized entries:
\[
\mathrm{SC}_{\mathrm{map}}
=
\mathrm{Cor}\!\left(\mathrm{vec}(M_{\mathrm{model}}),\,
                     \mathrm{vec}(M_{\mathrm{simple}})\right),
\]
where $\mathrm{Cor}(\cdot,\cdot)$ denotes the Pearson correlation
coefficient throughout this paper.
High values indicate that the estimated feedback-enhanced mapping remains
close to the direct-effect structure learned by the feedforward
initialization.
Accordingly, SC$_{\mathrm{map}}$ should be interpreted as a
\emph{continuity diagnostic relative to the feedforward baseline}, not as
an external goodness-of-fit statistic in the SEM sense.

\subsection{Second-moment fidelity}\label{subsec:SCcov}

Equilibrium predictions $\widehat{Y}_1=M_{\mathrm{model}}Y_2$ induce a
second-moment structure among endogenous variables given by
\[
S_{\mathrm{model}}
=\widehat{Y}_1\widehat{Y}_1^\top
=M_{\mathrm{model}}S_{Y_2}M_{\mathrm{model}}^\top,
\qquad S_{Y_2}=Y_2Y_2^\top .
\]
This matrix summarizes the dependence pattern implied by the equilibrium
input--output mapping. We compare it with the empirical second-moment
matrix $S_{\mathrm{sample}}=Y_1Y_1^\top$ via
\[
\mathrm{SC}_{\mathrm{cov}}
=
\mathrm{Cor}\!\left(\mathrm{vec}(S_{\mathrm{model}}),\,
                     \mathrm{vec}(S_{\mathrm{sample}})\right),
\]
which evaluates whether the equilibrium mapping reproduces the observed
second-moment structure among endogenous variables.

Throughout this paper, second-moment matrices are treated as
covariance-like summaries of dependence. All variables are pre-processed
to a common scale by min-max rescaling
$(x-\min)/(\max-\min)$. The model itself only requires that
$(Y_1,Y_2)$ be non-negative, so a strict $[0,1]$ range is not
formally necessary. We nonetheless recommend range alignment as the
default preprocessing for two estimator-fairness reasons:
(i) the Frobenius reconstruction loss
$\|Y_1 - X(\Theta_1 Y_1+\Theta_2 Y_2)\|_F^2$ is summed across all
variables of $Y_1$ with equal weight, so without range alignment
variables with larger numerical ranges would dominate the loss;
(ii) the $\ell_1$ regularization coefficients $(\lambda_1,\lambda_2)$
act uniformly across all variables of $Y_1$ (resp.\ $Y_2$), so
range disparities translate into uneven effective shrinkage across
variables. Aligning the per-variable range thus prevents any single
variable from being structurally favored or over-shrunk by the
loss/penalty geometry alone, while also improving the interpretability
of the structural correlations introduced below. The robustness
caveats of min-max rescaling are discussed in
Section~\ref{sec6}.

\subsection{Feedback strength}\label{subsec:feedback_metrics}

Latent feedback is summarized by two quantities:
the spectral radius $\rho(X\Theta_1)$ and the amplification ratio
$\mathrm{AR}$ defined in Section~\ref{subsec:leontief}.  
The former captures the intrinsic strength of the latent feedback loop,
while AR compares total equilibrium effects with direct effects.
Here, latent feedback should be interpreted as a parsimonious structural
representation of residual dependence among endogenous variables not explained by observed
exogenous variables, under the assumption that the observed endogenous
variables reflect equilibrium outcomes rather than transient dynamics.

Sampling uncertainty for $\rho(X\Theta_1)$ and AR, as well as for
the individual entries of $\Theta_1$ and $\Theta_2$, is assessed by a
single nonparametric \emph{X-fixed} bootstrap: we resample the $N$
columns of $(Y_1,Y_2)$ with replacement $B_{\mathrm{boot}}=1000$
times (the subscript distinguishes the resample count from the
latent score matrix $B=\Theta_1 Y_1+\Theta_2 Y_2$), and on each
resample $(Y_1^{(b)}, Y_2^{(b)})$ we re-estimate
$(\Theta_1,\Theta_2)$ \emph{conditional on the basis matrix $X$
held fixed at its original-data fit}. We then report 95\% percentile
intervals for $\rho(X\Theta_1)$ and AR (using the resampled
$\Theta_1^{(b)}$), and compute percentile intervals and one-sided
support rates for each entry of $\Theta_1$ and $\Theta_2$.
This is analogous to confirmatory SEM, in which the measurement
model is fixed first and structural coefficients are estimated
conditional on it.
Several implementation choices should be stated explicitly.
First, the regularization parameters
$(\lambda_X,\lambda_1,\lambda_2)$ are held fixed at the values
selected on the original sample by cross-validated MAE
(Section~\ref{subsec:crossvalidation}); the tuning step is not
repeated inside each bootstrap replicate, so the intervals are
conditional on the selected penalty level and slightly underestimate
variability relative to a fully nested double bootstrap that re-tunes
the penalties inside each replicate.
Second, fixing $X$ across replicates serves a dual purpose.
\emph{(i)~Factor-membership stability for $\Theta_2$ inference.}
When $X$ is re-estimated per resample, the multiplicative updates
can not only permute the latent factor order but also reshuffle
which $Y_1$ outcomes are bundled into each factor
(``factor-membership instability'', especially acute for small-$N$
resamples such as the Mississippi application). Since each
$\Theta_2$ entry is interpreted as the effect of a $Y_2$ covariate
on a \emph{specific} latent factor, any change in factor membership
cascades into large, label-driven shifts in the apparent significance
of $\Theta_2$ entries, confounding sampling variability with
definitional drift of the latent factors themselves. Fixing $X$ at
the original-data fit removes this artifact, so the reported support
rates reflect sampling uncertainty in $\Theta_1$ and $\Theta_2$
\emph{given a stable factor definition} rather than confound it
with re-bundling of $Y_1$.
\emph{(ii)~Avoidance of per-replicate SVD instability.} It also
avoids the spectral instability that resampling-with-replacement can
introduce into a per-replicate SVD initialization of $X$. In a
bootstrap sample of size $N$ drawn with replacement, the probability
that a given observation is never selected is
$(1-1/N)^{N}\!\to\!e^{-1}\!\approx\!0.368$, so an expected
$\sim\!37\%$ of the bootstrap slots are duplicates of observations
already drawn. The trade-off is that the resulting
$\rho(X\Theta_1)$ and AR intervals are conditional on the
data-driven measurement model and therefore underestimate
variability that a fully unconditional bootstrap (with $X$
re-estimated per replicate) would expose.
Third, replicates for which the multiplicative updates failed to
produce a stable equilibrium ($\rho(X\Theta_1)\ge 1$, or
non-convergence within $t_{\max}$) are discarded prior to percentile
computation; the empirical discard rate was $0\%$ for HS39, $0\%$ for
the LA data, and below $3\%$ for the Mississippi data, so the
percentile intervals are minimally affected by truncation.

\paragraph*{Inference for individual path coefficients.}
Because the parameter space is restricted to non-negative entries
($\theta\ge 0$), tests of $H_0\!:\theta=0$ have a structurally
\emph{one-sided} alternative $H_1\!:\theta>0$. We report the
bootstrap support rate
$\Pr_{\mathrm{boot}}(\hat\theta^{(b)}>0.01)$
on each entry of $\Theta_1$ and $\Theta_2$, where $0.01$ matches
the two-decimal display threshold used throughout the manuscript;
$*$, $**$, and $***$ denote support rates exceeding $0.95$, $0.99$,
and $0.999$, respectively (one-sided bootstrap $p$-value
$\le 0.05$, $0.01$, $0.001$).
These markers appear on the NMF-FFB columns of
Tables~\ref{tab:hs39_struct_compare}, \ref{tab:la_struct_compare},
and~\ref{tab:ms_struct_compare} for $\Theta_2$, and on the latent-feedback edges in
Figs.~\ref{fig:hs39_path}--\ref{fig:mshealth_path} for $\Theta_1$
(no markers are placed on the basis $X$ edges of the path diagrams,
since $X$ is held fixed across replicates by construction).
For all $60$ entries of $\Theta_2$ across the three applications
($9$ for HS39, $18$ for LA, $33$ for Mississippi),
the original-data $\hat\theta_2$ point estimate lies inside the
corresponding 95\% percentile bootstrap CI, providing a sanity check
that the X-fixed percentile bootstrap behaves consistently with
the original-data fit.
These intervals should therefore be read as uncertainty bands for the
stable-fitted regime under fixed penalties, not as unconditional
sampling distributions over all possible refits.

\subsection{Simulation summary}\label{subsec:simulation}

To verify that the proposed metrics behave coherently when the true
system is known, we conducted Monte Carlo experiments under controlled
data-generating conditions.  We varied both the noise level
$\sigma\in\{0,\,0.10\}$ and the strength of latent feedback
$\rho_{\mathrm{true}}\in\{0,\,0.2\}$ in order to assess the sensitivity
of each metric to these key structural features.
Throughout this section, $\sigma$ denotes the standard deviation of the
Gaussian noise added (and truncated at zero) to the entries of the
generated $Y_1$ before fitting, while
$\rho_{\mathrm{true}} = \rho(X_{\mathrm{true}}\Theta_{1,\mathrm{true}})$
is the spectral radius of the true latent-feedback operator used to
generate the data, so that $\rho_{\mathrm{true}}=0$ corresponds to a
purely feedforward ground truth.
In each setting,
synthetic data were generated from fixed non-negative parameters
$(X_{\mathrm{true}},\Theta_{1,\mathrm{true}},\Theta_{2,\mathrm{true}})$,
and samples of size $N\in\{50,200\}$ were drawn.  Models were then fitted
using the estimation procedure described in Section~\ref{sec3}, and
results were aggregated over $R=500$ replications.
Specifically, all three simulation tables of \S\ref{subsec:simulation}
and \S\ref{subsec:comparison} fix the dimensions at $P_1{=}9$, $P_2{=}3$,
and the true latent rank at $Q_{\mathrm{true}}=3$, mirroring the
HS39 application of \S\ref{subsec:hs39} so that the simulation
diagnostics are directly comparable to the empirical analysis.
Each column of $X_{\mathrm{true}}$ was generated by drawing entries
i.i.d.\ from $\mathrm{Exponential}(1)$ and then normalizing the column
to sum to one (column-stochastic normalization, equivalent in
distribution to a uniform Dirichlet).
The matrix $\Theta_{1,\mathrm{true}}$ was constructed as a sparse
template in which each row receives two non-zero entries drawn from
$\mathrm{Uniform}(0.2,\,1)$ at randomly chosen column positions, and
this template was then uniformly rescaled so that
$\rho(X_{\mathrm{true}}\Theta_{1,\mathrm{true}})=\rho_{\mathrm{true}}$
exactly.
The matrix $\Theta_{2,\mathrm{true}}$ was likewise constructed
column-wise: for each column, one or two row positions were drawn
uniformly at random and assigned values from $\mathrm{Uniform}(0.5,\,2)$,
with the remaining entries set to zero.
Exogenous inputs $Y_2$ were drawn elementwise from $\mathrm{Uniform}(0,1)$,
and the equilibrium signal
\(
Y_1^{(0)} = (I - X_{\mathrm{true}}\Theta_{1,\mathrm{true}})^{-1}
            X_{\mathrm{true}}\Theta_{2,\mathrm{true}}Y_2
\)
was perturbed by additive Gaussian noise with standard deviation
$\sigma$, truncated at zero to preserve non-negativity.

Table~\ref{tab:sim00} reports the full Monte Carlo results across two
noise levels $\sigma\in\{0,\,0.10\}$, two feedback values
$\rho_{\mathrm{true}}\in\{0,\,0.2\}$, and two sample sizes
$N\in\{50,200\}$ (8 conditions, $R=500$ replications each).

\begin{table}[htbp]
\centering
\small
\caption{Monte Carlo simulation results under the fixed-penalty
convention used in the main study ($R=500$ replications per condition;
$P_1{=}9$, $P_2{=}3$, $Q{=}3$ matching the HS39 dimensions of
\S\ref{subsec:hs39}). Values are means across replications.
Two noise levels $\sigma\in\{0,0.10\}$ are reported for each
combination of true feedback strength
$\rho_{\mathrm{true}}\in\{0,0.2\}$ and sample size $N\in\{50,200\}$
(8 conditions total). Under this fixed-penalty convention, $\hat\rho$
and $\widehat{\mathrm{AR}}$ should be read as exploratory within-regime
diagnostics, not as scale-free quantities for cross-$N$ or
cross-dataset comparison.}
\label{tab:sim00}
\begin{tabular}{lccccc}
\hline
Condition & $\hat\rho$ & $\widehat{\mathrm{AR}}$
& SC$_{\mathrm{map}}$ & SC$_{\mathrm{cov}}$ & MAE \\
\hline
$\rho_{\mathrm{true}}{=}0$,\; $\sigma{=}0$,\; $N{=}50$     & 0.014 & 1.01 & 0.985 & 1.000 & 0.009 \\
$\rho_{\mathrm{true}}{=}0$,\; $\sigma{=}0$,\; $N{=}200$    & 0.122 & 1.12 & 0.989 & 1.000 & 0.007 \\
$\rho_{\mathrm{true}}{=}0$,\; $\sigma{=}0.10$,\; $N{=}50$  & 0.033 & 1.03 & 0.978 & 0.999 & 0.075 \\
$\rho_{\mathrm{true}}{=}0$,\; $\sigma{=}0.10$,\; $N{=}200$ & 0.455 & 1.51 & 0.976 & 0.999 & 0.077 \\
\hline
$\rho_{\mathrm{true}}{=}0.2$,\; $\sigma{=}0$,\; $N{=}50$     & 0.032 & 1.03 & 0.986 & 1.000 & 0.010 \\
$\rho_{\mathrm{true}}{=}0.2$,\; $\sigma{=}0$,\; $N{=}200$    & 0.180 & 1.26 & 0.992 & 1.000 & 0.008 \\
$\rho_{\mathrm{true}}{=}0.2$,\; $\sigma{=}0.10$,\; $N{=}50$  & 0.048 & 1.05 & 0.980 & 0.999 & 0.077 \\
$\rho_{\mathrm{true}}{=}0.2$,\; $\sigma{=}0.10$,\; $N{=}200$ & 0.530 & 2.00 & 0.973 & 0.999 & 0.079 \\
\hline
\end{tabular}
\end{table}

Accordingly, Table~\ref{tab:sim00} should be read primarily as a
descriptive check of how the fixed-penalty procedure behaves across the
simulation grid, not as a fairness-adjusted comparison of absolute
$\hat\rho$ or AR magnitude across different sample sizes.

Three patterns emerge.  First, MAE tracks the injected noise level
directly: mean MAE is about $0.009$ at $\sigma{=}0$ (range
$0.007$--$0.009$ across the four rows) and about $0.077$ at
$\sigma{=}0.10$ (range $0.075$--$0.077$), with only small variation
over $N$ and $\rho_{\mathrm{true}}$.  This confirms that MAE is
dominated by sampling noise rather than by structural misspecification.

Second, $\hat\rho$ increases with $\sigma$, with $N$, and with
$\rho_{\mathrm{true}}$; the most pronounced case is the
$\rho_{\mathrm{true}}{=}0,\sigma{=}0.10$ row, where $\hat\rho$ rises
from $0.033$ at $N{=}50$ to $0.455$ at $N{=}200$. Under the fixed-penalty convention of
Table~\ref{tab:sim00}, however, $\hat\rho$ should be read as an
\emph{exploratory feedback diagnostic}, not as a scale-free estimator of
$\rho_{\mathrm{true}}$ and not as stand-alone evidence that genuine
feedback is present.

Two mechanisms drive the sensitivity to $N$. (i) Non-negativity
produces a positive finite-sample bias in $\hat\rho$: small positive
$\Theta_1$ entries arising from sampling variation cannot shrink
below zero, which explains the nonzero $\hat\rho$ in the
$\rho_{\mathrm{true}}=0$ rows ($\hat\rho{=}0.014$ at
$N{=}50,\sigma{=}0$). (ii) Because the reconstruction error grows
as $O(N)$ while $\lambda_X$ is held fixed across conditions in
Table~\ref{tab:sim00}, the effective orthogonality strength decays
as $1/N$, contributing a systematic size-dependent inflation of
$\hat\rho$ visible in the rise from $\hat\rho{=}0.014$ at $N{=}50$
to $\hat\rho{=}0.122$ at $N{=}200$ (both $\rho_{\mathrm{true}}{=}0$,
$\sigma{=}0$), and from $\hat\rho{=}0.033$ to $\hat\rho{=}0.455$
under noise. Section~\ref{subsubsec:Nscale} shows that scaling all
three penalties by $N/N_0$ greatly reduces this second component (e.g.\
$0.122 \to 0.010$ and $0.455 \to 0.041$ at $N{=}200$ under the
$N$-scaled convention); absolute magnitudes of $\hat\rho$ and AR
should accordingly be compared across $N$ only under that
$N$-scaled convention.

Third, $\widehat{\mathrm{AR}}$ increases with $\hat\rho$,
consistent with the geometric amplification bound of
Section~\ref{subsec:arbounds}: larger $\hat\rho$ produces larger
column sums in $X\Theta_1$ and hence higher amplification through
the Leontief inverse $(I - X\Theta_1)^{-1}$.  Across the eight
conditions of Table~\ref{tab:sim00}, $\widehat{\mathrm{AR}}$
ranges from $\approx 1.01$ (well-controlled) to $\approx 2.00$ at
$\sigma{=}0.10, N{=}200, \rho_{\mathrm{true}}{=}0.2$, in the same
direction as the unregularized regime of NMF-FFB$_{\mathrm{unreg}}$
(Section~\ref{subsec:comparison}) but well short of the
divergent values ($\widehat{\mathrm{AR}}\approx 200$--$350$)
observed there.

Across all eight conditions, SC$_{\mathrm{map}}\ge 0.97$ and
SC$_{\mathrm{cov}}\ge 0.999$, confirming that the latent basis $X$
and its induced covariance structure are recovered faithfully even
under moderate noise.  The proposed metrics therefore respond
coherently to controlled changes in the data-generating process
while preserving structural fit.

\subsection{Comparison with ablated and benchmark methods}
\label{subsec:comparison}

To clarify the role of each NMF-FFB component, we compare five methods:
\begin{itemize}
\item[(i)] \textbf{NMF-FFB} --- the full proposed method.
\item[(ii)] \textbf{NMF-FF} --- the ablation $\Theta_1 = 0$
      (feedforward only; no latent feedback).
\item[(iii)] \textbf{NMF-FFB$_{\mathrm{unreg}}$} --- the unregularized
      variant $\lambda_X=\lambda_1=\lambda_2=0$ (no orthogonality,
      no sparsity).
\item[(iv)] \textbf{OLS-MIMIC} --- the saturated linear
      Multiple-Indicators-Multiple-Causes (MIMIC) benchmark
      $Y_1 = \Xi Y_2 + E$ (with $\Xi \in \mathbb{R}^{P_1\times P_2}$)
      fitted by ordinary least squares, without non-negativity and
      without latent structure. This serves as a uniform signed
      baseline that can be applied consistently across all three
      datasets.
\item[(v)] \textbf{Standard SEM} --- a literal standard SEM
      \emph{without non-negativity constraints}, fitted by maximum
      likelihood; throughout we use the \texttt{lavaan} R package
      \citep{lavaan} for the implementation, and the corresponding
      table row is labelled \emph{Standard SEM} in
      Table~\ref{tab:real_comp}. We use for each dataset the same
      measurement-and-structural specification reported in the
      corresponding subsection of
      Section~\ref{sec5}: a three-factor MIMIC for HS39
      (canonical groupings $x_1$--$x_3$, $x_4$--$x_6$, $x_7$--$x_9$,
      each latent factor regressed on \texttt{age.rev},
      \texttt{sex.2}, \texttt{school.GW}); a hybrid path-analytic
      model for LA (one cardiovascular latent factor measured by
      \texttt{tmort} and \texttt{cmort} plus \texttt{rmort} as a
      directly observed outcome, both regressed on nine
      meteorological/pollution covariates); and a three-factor MIMIC
      for Mississippi using the NMF-FFB-discovered groupings
      (early-mortality, distress, and despair indicators), each
      latent factor regressed on the eleven social-determinant
      covariates.
\end{itemize}
The simulation comparison uses a data-generating process matching the
HS39 dimensions ($P_1{=}9$, $P_2{=}3$, $Q{=}3$) at the small-sample
size $N{=}50$ that mirrors the Mississippi application, with
$R=500$ replications per condition (Table~\ref{tab:sim_comp}); for
method (v), the lavaan factor structure is determined per replicate
by varimax-rotated PCA so that the SEM specification adapts to the
sampled data; convergence failures are flagged in the corresponding
table notes.
Real-data comparisons use the three applications of \S\ref{sec5}
(HS39, LA, Mississippi; see \S\ref{sec1} for abbreviations) with
NMF-FFB hyperparameters fixed at the values selected for the
corresponding application.

The Standard SEM benchmark of method (v) (implemented via the
\texttt{lavaan} package) exhibits dataset-specific behavior. HS39 admits the canonical three-factor
model and converges with moderate fit
($\chi^2(42)=149.1$, CFI $=0.893$, TLI $=0.840$, RMSEA $=0.092$,
SRMR $=0.066$; robust MLR gives the same qualitative conclusion).
LA pollution requires the hybrid path-analytic specification listed
above to mirror the NMF-FFB grouping, since $P_1=3$ endogenous
indicators do not support two latent factors with multiple indicators
each; even then absolute fit is poor (TLI $=0.692$, RMSEA $=0.193$).
Mississippi sits at the boundary of empirical identifiability
($\sim$$50$ free ML parameters against $N=35$): \texttt{lavaan} with
its default NLMINB optimizer fails to converge, and switching to BFGS
produces a statistically degenerate fixed point (standardized
coefficients $\!>\!1$ in absolute value, standard errors up to
$\sim$$2{,}000$, $\mathrm{CFI}\!\approx\!0.65$); for the prediction-
level comparison of Table~\ref{tab:real_comp} we report the lavaan
BFGS fit as a faithful disclosure of this small-$N$ convergence
failure.\footnote{For Mississippi we additionally attempted
\texttt{OpenMx} \citep{openmx} (converges numerically but with
unsalvageable fit, $\mathrm{CFI}\!\approx\!0.77$,
$\mathrm{RMSEA}\!\approx\!0.22$); the \texttt{sem} package
\citep{fox2006} (fails outright on a non-invertible model-implied
covariance matrix); and a model-implied instrumental-variables 2SLS
estimator \citep[\texttt{MIIVsem};][]{bollen1996,bollen2021}, which
yields the only structural-coefficient SEM benchmark with usable
standard errors and is reported in Section~\ref{subsec:mshealth}.}

Table~\ref{tab:real_comp} reports the real-data comparison.
The unregularized variant NMF-FFB$_{\mathrm{unreg}}$ fails in all three
datasets: the spectral radius $\rho(X\Theta_1)$ saturates at unity and
the amplification ratio explodes (AR $=629$ for HS39, $3643$ for LA,
and $170$ for Mississippi), while
the structural correlation SC$_{\mathrm{map}}$ collapses and even
becomes negative in the LA case.  This demonstrates that the
orthogonality and sparsity penalties are essential for producing stable
feedback estimates; without them, the Leontief inverse is
numerically unstable.

The feedforward ablation NMF-FF attains MAE nearly identical to
the full NMF-FFB on all three datasets, reflecting that predictive fit
alone does not require feedback.  However, NMF-FF is forced to
return $\rho=0$ by construction, so it cannot distinguish systems with
genuine feedback (HS39, $\rho\approx 0.36$) from those without (LA,
$\rho\approx 0.002$). The full NMF-FFB distinguishes these regimes
and provides an interpretable diagnostic of feedback strength.

OLS-MIMIC achieves lower MAE than NMF-FFB on all three datasets, as
expected from its saturated linear form with $P_1 P_2$ free parameters
and no constraints; the Standard SEM row can be still lower in HS39 but
is dataset-specific and less uniformly admissible across the three
applications.
However, it provides no non-negative parts-based decomposition, no
latent component interpretation, and no feedback diagnosis: the
estimated $B$ contains signed coefficients that do not admit an IO
reading, and the method offers no analogue of the Leontief inverse.
The price for predictive accuracy is therefore the loss of structural
interpretability that motivates NMF-FFB.

\paragraph*{Information asymmetry in the Standard SEM benchmark.}
The five-method comparison in Table~\ref{tab:real_comp} is
methodologically asymmetric in a way worth making explicit. Standard
SEM requires the measurement structure---which observed indicators
load on which latent factor---to be specified before fitting; in
our three applications this specification is supplied externally
(canonical $x_1$--$x_3$, $x_4$--$x_6$, $x_7$--$x_9$ for HS39; a
hybrid path-analytic specification adopted for LA so that the SEM
can converge at all; and, for Mississippi, the very grouping
discovered by NMF-FFB itself). NMF-FFB, by contrast, is given only
the latent rank $Q$ and recovers the measurement structure (the
column-stochastic basis $X$) jointly with the structural coefficients
$(\Theta_1,\Theta_2)$. The comparison therefore evaluates Standard
SEM with measurement information pre-supplied, whereas NMF-FFB must
learn that measurement structure from the data; a symmetric benchmark would
require an exploratory factor analysis or comparable
dimensionality-reduction step before the confirmatory SEM, yielding a
two-stage procedure whose combined uncertainty is rarely propagated.
A fair reading of Table~\ref{tab:real_comp} is therefore that
the NMF-FFB MAE stays within
$0.04$, $0.01$, and $0.01$ of the Standard SEM benchmark on HS39, LA,
and Mississippi respectively; in these applications, data-driven
measurement recovery incurs only a modest absolute predictive cost
while retaining the latent-feedback and parts-based outputs that
motivate NMF-FFB.

\begin{table}[htbp]
\centering\small
\caption{Real-data comparison of NMF-FFB against ablations
(NMF-FF, NMF-FFB$_{\mathrm{unreg}}$) and two unconstrained-SEM
benchmarks (OLS-MIMIC; Standard SEM, fitted by maximum likelihood
via the \texttt{lavaan} R package \citep{lavaan}, with the
dataset-specific specifications detailed in method (v) of
\S\ref{subsec:comparison}).
NMF-FFB hyperparameters are fixed at the values selected in
Section~\ref{sec5}.
Dashes (---) mark metrics undefined for that method
($\rho$ and AR require non-negative latent feedback;
SC$_{\mathrm{map}}$ requires a non-negative latent basis).
$^\dagger$For Mississippi the lavaan ML optimizer fails to converge
($N{=}35$ vs $\sim$$50$ free parameters); BFGS is used and the SEM
fit is poor (CFI${\approx}0.65$, RMSEA${\approx}0.22$;
cf.\ \S\ref{subsec:mshealth}).}
\label{tab:real_comp}
\begin{tabular}{llccccc}
\hline
Dataset & Method & $\rho(X\Theta_1)$ & AR &
SC$_{\mathrm{map}}$ & SC$_{\mathrm{cov}}$ & MAE \\
\hline
HS39
 & NMF-FFB                & 0.357 &  1.36   & 0.999 & 0.980 & 0.181 \\
 & NMF-FF            & 0.000 &  1.00   & 1.000 & 0.978 & 0.181 \\
 & NMF-FFB$_{\mathrm{unreg}}$& 1.002 & 629     & 0.669 & 0.882 & 2.971 \\
 & OLS-MIMIC             & ---   & ---     & ---   & 0.979 & 0.180 \\
 & Standard SEM          & ---   & ---     & ---   & 0.980 & 0.142 \\
\hline
LA
 & NMF-FFB                & 0.002 &  1.00   & 0.882 & 0.983 & 0.095 \\
 & NMF-FF            & 0.000 &  1.00   & 1.000 & 0.984 & 0.094 \\
 & NMF-FFB$_{\mathrm{unreg}}$& 1.000 & 3643    &$-0.675$& 0.982 & 3.764 \\
 & OLS-MIMIC             & ---   & ---     & ---   & 0.988 & 0.088 \\
 & Standard SEM          & ---   & ---     & ---   & 0.988 & 0.088 \\
\hline
Mississippi
 & NMF-FFB                & 0.446 &  1.02   & 0.950 & 0.989 & 0.117 \\
 & NMF-FF            & 0.000 &  1.00   & 1.000 & 0.989 & 0.118 \\
 & NMF-FFB$_{\mathrm{unreg}}$& 1.003 & 170     & 0.009 & 0.926 & 1.718 \\
 & OLS-MIMIC             & ---   & ---     & ---   & 0.997 & 0.092 \\
 & Standard SEM$^\dagger$& ---   & ---     & ---   & 0.995 & 0.106 \\
\hline
\end{tabular}
\end{table}

The ablations make the trade-offs visible: removing feedback
(NMF-FF) sacrifices diagnostic information without improving fit;
removing regularization (NMF-FFB$_{\mathrm{unreg}}$) destabilizes the
Leontief inverse; OLS-MIMIC attains marginally lower MAE while
sacrificing parts-based and equilibrium-IO interpretation. The
Standard SEM benchmark attains MAE in the same range as NMF-FFB
and OLS-MIMIC on HS39 and LA (with a $0.04$ HS39 gap due to
lavaan's intercept term in its mean structure), but on the
small-$N$ Mississippi case lavaan's ML optimization fails outright
(only the BFGS fallback converges, with $\mathrm{CFI}{\approx}0.65$).
The methodological payoff of NMF-FFB is therefore not lower MAE but
the additional structural outputs ($\Theta_1$, the Leontief inverse
$(I-X\Theta_1)^{-1}$, AR, and parts-based interpretability of $X$)
that standard SEM does not produce.

The effect of removing regularization is visible not only at the
scalar level (Table~\ref{tab:real_comp}: $\widehat{\mathrm{AR}}$ rising
from $1.36$ to $629$ on HS39 under
NMF-FFB$_{\mathrm{unreg}}$) but also at the matrix level. On HS39, the
off-ratio of $X^\top X$ rises from $6.1\times 10^{-3}$ at
$\lambda_X{=}100$ to $4.4\times 10^{-2}$ at $\lambda_X{=}0$ (with
$\lambda_1{=}\lambda_2{=}0$ as well), the maximum cosine between basis
columns rises from $0.079$ to $0.179$, and $\hat\rho(X\Theta_1)$ is
driven to the stability boundary ($\hat\rho\approx 1.00$). In other
words, the fitted $X$ loses its column separation and the path
coefficients $\Theta_1, \Theta_2$ become dense, so the parts-based
factor structure and the equilibrium Leontief operator that give
NMF-FFB its interpretation collapse simultaneously with the scalar
diagnostics.

The simulation comparison (summarized in
Table~\ref{tab:sim_comp}) confirms these conclusions under the
controlled data-generating process.

\begin{table}[htbp]
\centering
\small
\caption{Simulation comparison (mean over $R=500$ replications;
$P_1{=}9$, $P_2{=}3$, $Q{=}3$ matching the HS39 dimensions of
\S\ref{subsec:hs39}; $N{=}50$, a small-sample stress test that mirrors
the Mississippi application of \S\ref{subsec:mshealth}).
Conditions vary the true feedback strength
$\rho_{\mathrm{true}}\in\{0, 0.2\}$ and noise level
$\sigma\in\{0, 0.10\}$.
OLS-MIMIC and Standard SEM are methods (iv) and (v) of
\S\ref{subsec:comparison}.
SC$_{\mathrm{map}}^{\mathrm{true}}$ is the correlation between the
estimated equilibrium mapping $M_{\mathrm{model}}$ and the true
$M_{\mathrm{true}}$.
Dashes (---) mark metrics undefined for a method
($\rho$ and AR for OLS-MIMIC and Standard SEM).
$^\dagger$Standard SEM (lavaan ML) is undefined at $\sigma{=}0$:
zero residual variance violates the ML likelihood and all
replicates fail to converge.}
\label{tab:sim_comp}
\begin{tabular}{llcccc}
\hline
Condition & Method & $\hat\rho$ & $\widehat{\mathrm{AR}}$ &
SC$_{\mathrm{map}}^{\mathrm{true}}$ & MAE \\
\hline
$\rho_{\mathrm{true}}{=}0$, $\sigma{=}0$
 & NMF-FFB                  & 0.014 &   1.01 & 0.984 & 0.009 \\
 & NMF-FF                   & 0.000 &   1.00 & 1.000 & 0.001 \\
 & NMF-FFB$_{\mathrm{unreg}}$& 0.805 &   5.63 & 0.979 & 0.008 \\
 & OLS-MIMIC                & ---   & ---    & 1.000 & 0.000 \\
 & Standard SEM$^\dagger$   & ---   & ---    & ---   & ---   \\
\hline
$\rho_{\mathrm{true}}{=}0$, $\sigma{=}0.10$
 & NMF-FFB                  & 0.032 &   1.02 & 0.959 & 0.075 \\
 & NMF-FF                   & 0.000 &   1.00 & 0.972 & 0.074 \\
 & NMF-FFB$_{\mathrm{unreg}}$& 0.988 & 348    & 0.521 & 0.360 \\
 & OLS-MIMIC                & ---   & ---    & 0.968 & 0.073 \\
 & Standard SEM             & ---   & ---    & 0.942 & 0.075 \\
\hline
$\rho_{\mathrm{true}}{=}0.2$, $\sigma{=}0$
 & NMF-FFB                  & 0.032 &   1.03 & 0.985 & 0.010 \\
 & NMF-FF                   & 0.000 &   1.00 & 1.000 & 0.001 \\
 & NMF-FFB$_{\mathrm{unreg}}$& 0.809 &   5.93 & 0.986 & 0.007 \\
 & OLS-MIMIC                & ---   & ---    & 1.000 & 0.000 \\
 & Standard SEM$^\dagger$   & ---   & ---    & ---   & ---   \\
\hline
$\rho_{\mathrm{true}}{=}0.2$, $\sigma{=}0.10$
 & NMF-FFB                  & 0.057 &   1.07 & 0.961 & 0.077 \\
 & NMF-FF                   & 0.000 &   1.00 & 0.978 & 0.075 \\
 & NMF-FFB$_{\mathrm{unreg}}$& 0.984 & 231    & 0.620 & 0.252 \\
 & OLS-MIMIC                & ---   & ---    & 0.975 & 0.075 \\
 & Standard SEM             & ---   & ---    & 0.954 & 0.077 \\
\hline
\end{tabular}
\end{table}

Three patterns emerge consistently across the conditions.
(i)~\textbf{NMF-FFB$_{\mathrm{unreg}}$ destabilizes under noise.}
The unregularized variant drives $\hat\rho$ toward the stability
boundary ($\hat\rho\ge 0.80$ in every condition) and the amplification
ratio $\widehat{\mathrm{AR}}$ explodes from about $5$--$6$ at
$\sigma{=}0$ to $230$--$350$ at $\sigma{=}0.10$, with the equilibrium
fit $\mathrm{SC}_{\mathrm{map}}^{\mathrm{true}}$ collapsing from
$\approx 0.98$ to $0.52$--$0.62$ and MAE inflating to $0.25$--$0.36$.
This confirms that the Leontief inverse becomes numerically unstable
without the regularizers $\lambda_X$, $\lambda_1$, $\lambda_2$.
(ii)~\textbf{NMF-FF, OLS-MIMIC, and Standard SEM achieve comparable
MAE to NMF-FFB at $\sigma{=}0.10$} (all in the $0.073$--$0.077$ range),
matching the pattern observed on the real datasets in
Table~\ref{tab:real_comp}: NMF-FFB does not pay a meaningful
predictive cost for its non-negative parts-based decomposition.
(iii)~\textbf{NMF-FFB's $\hat\rho$ shows substantial small-$N$
shrinkage and is correspondingly read as an ordinal diagnostic.}
At $\rho_{\mathrm{true}}{=}0$, NMF-FFB shows a small positive
non-negativity floor ($\hat\rho \in \{0.014, 0.032\}$): sampling
fluctuations in $\Theta_1$ cannot shrink below zero. At
$\rho_{\mathrm{true}}{=}0.2$, $\hat\rho \in \{0.032, 0.057\}$
--- substantially \emph{below} $\rho_{\mathrm{true}}$. This
small-$N$ shrinkage is consistent with the corresponding
$N{=}50$ rows of Table~\ref{tab:sim00}
($\rho_{\mathrm{true}}{=}0.2,\sigma{=}0$:
$\hat\rho{=}0.032$; $\sigma{=}0.10$: $\hat\rho{=}0.048$): the L1
regularization on $\Theta_1$ pulls $\hat\rho$ toward the
non-negativity floor when the parameter-to-sample ratio is
sizeable ($Q\!\times\!P_1 / N = 27/50$ here). Reinforcing our
methodological stance, $\hat\rho$ should be read as an ordinal
feedback diagnostic rather than a scale-free point estimator
(\S\ref{subsec:feedback_metrics}, \S\ref{subsubsec:Nscale}); the
bootstrap percentile interval reported alongside the point
estimate in \S\ref{sec5} provides the appropriate calibration.

\subsubsection*{Validation of the $N$-scaled penalty convention}
\label{subsubsec:Nscale}

Because the reconstruction loss $\|Y_1-XB\|_F^{2}$ grows as $O(N)$
while $\|\Theta_1\|_1$ and $\|\Theta_2\|_1$ are $O(1)$, any
\emph{fixed} choice of $(\lambda_X,\lambda_1,\lambda_2)$ implicitly
imposes effective regularization of order $1/N$, so nominally identical
penalties become progressively weaker as $N$ grows. We therefore
adopted in Section~\ref{subsec:regularization} the $N$-scaled convention
$(\lambda_X,\lambda_1,\lambda_2)\mapsto (N/N_0)\,(\lambda_X,\lambda_1,\lambda_2)$
with reference $N_0=50$ as a \emph{design choice} that keeps effective
regularization comparable across sample sizes, rather than as a
post-hoc correction. This subsection reports empirical validation of
that convention. We recompute NMF-FFB at the larger sample size
$N{=}200$ under the $N$-scaled convention with the same DGP and
conditions as the $N{=}200$ rows of Table~\ref{tab:sim00}
($\sigma\in\{0,0.10\}$, $\rho_{\mathrm{true}}\in\{0,0.2\}$,
$R=500$ replications per cell), and report the resulting metrics
side-by-side with the corresponding fixed-penalty entries of
Table~\ref{tab:sim00} in Table~\ref{tab:Nscale}.

\begin{table}[htbp]
\centering\small
\caption{Side-by-side comparison of fixed-penalty (re-displayed from
Table~\ref{tab:sim00}, $N{=}200$ rows) versus $N$-scaled penalty
(reference $N_0{=}50$) on the same DGP at $N{=}200$ for the four
$(\rho_{\mathrm{true}},\sigma)$ conditions; same dimensions as
Table~\ref{tab:sim00} ($P_1{=}9$, $P_2{=}3$, $Q{=}3$). Values are
means over $R{=}500$ replications. The $N$-scaled convention greatly
reduces the apparent sample-size inflation in $\hat\rho$ and
$\widehat{\mathrm{AR}}$: at $\rho_{\mathrm{true}}{=}0, \sigma{=}0.10$,
fixed-penalty $\hat\rho$ at $N{=}200$ is $0.455$ whereas the
$N$-scaled value drops to $0.041$, an order-of-magnitude reduction
that returns $\hat\rho$ to the bias floor seen at $N{=}50$ in
Table~\ref{tab:sim00}.}
\label{tab:Nscale}
\begin{tabular*}{\textwidth}{@{\extracolsep{\fill}}cclccccc@{}}
\hline
$\rho_{\mathrm{true}}$ & $\sigma$ & Penalty &
$\hat\rho$ & $\widehat{\mathrm{AR}}$ &
SC$_{\mathrm{map}}$ & SC$_{\mathrm{cov}}$ & MAE \\
\hline
0.0 & 0.00 & Fixed       & 0.122 & 1.12 & 0.989 & 1.000 & 0.007 \\
0.0 & 0.00 & $N$-scaled  & 0.010 & 1.01 & 0.985 & 1.000 & 0.009 \\
0.0 & 0.10 & Fixed       & 0.455 & 1.51 & 0.976 & 0.999 & 0.077 \\
0.0 & 0.10 & $N$-scaled  & 0.041 & 1.05 & 0.983 & 0.999 & 0.077 \\
\hline
0.2 & 0.00 & Fixed       & 0.180 & 1.26 & 0.992 & 1.000 & 0.008 \\
0.2 & 0.00 & $N$-scaled  & 0.014 & 1.01 & 0.987 & 1.000 & 0.010 \\
0.2 & 0.10 & Fixed       & 0.530 & 2.00 & 0.973 & 0.999 & 0.079 \\
0.2 & 0.10 & $N$-scaled  & 0.071 & 1.09 & 0.983 & 0.999 & 0.079 \\
\hline
\end{tabular*}
\end{table}

Two conclusions follow. First, MAE and SC are essentially
unchanged between conventions (MAE stays in the $0.077$--$0.079$ range
under $\sigma{=}0.10$ and in the $0.009$--$0.010$ range under
$\sigma{=}0$ in both Fixed and $N$-scaled rows), so the $N$-scaling
correction does not materially alter predictive accuracy in this
experiment.
Second, the residual mean inflation under noise remains small under the
$N$-scaled convention but is not exactly zero. $\hat\rho$ should
accordingly be read as ordinal information about feedback strength
rather than as a scale-free point estimator. We recommend reporting $\hat\rho$ and
$\widehat{\mathrm{AR}}$ under the $N$-scaled convention for
cross-dataset comparisons, and under the fixed-$\lambda$ convention
used in Section~\ref{sec5} for within-dataset reading.

\subsection{Verification of the orthogonality penalty}
\label{subsec:ortho_check}

We next verify that the orthogonality penalty (the squared Frobenius
norm on the off-diagonal of $X^{\top}X$, introduced in
\S\ref{subsec:regularization}) actually achieves approximate
orthogonality (strictly, diagonality) of the latent basis columns.  The penalty term takes the
form
\[
  \tfrac{\lambda_X}{2}\sum_{i \ne j}\bigl(X^{\top}X\bigr)_{ij}^{2},
\]
so it drives the off-diagonal block of $X^{\top}X$ to zero while leaving
the diagonal $\mathrm{diag}(X^{\top}X)$ unconstrained.
To quantify how close the fitted basis is to diagonality we report
\[
  \text{off-ratio}
  \;=\;
  \frac{\|\mathrm{offdiag}(X^{\top}X)\|_{F}^{2}}
       {\|X^{\top}X\|_{F}^{2}}\in[0,1],
  \qquad
  \text{cos-max}
  \;=\;
  \max_{i \ne j}
  \frac{|(X^{\top}X)_{ij}|}{\sqrt{(X^{\top}X)_{ii}(X^{\top}X)_{jj}}},
\]
with off-ratio $=0$ and cos-max $=0$ corresponding to exact diagonality
(orthogonal columns of $X$).  We scan
$\lambda_X\in\{0,1,10,100\}$ using the same NNDSVDar-plus-feedforward
initialization protocol described in the caption of
Table~\ref{tab:ortho_real}; $\lambda_X=100$ is the value used for all results
in Sections~\ref{subsec:simulation}, \ref{subsec:comparison}, and
\ref{sec5}.

Table~\ref{tab:ortho_real} shows the result on the three real
datasets. At the default $\lambda_X{=}100$ the off-ratio is $0.6\%$
on HS39, $1.4\times 10^{-7}$ on LA, and $0.06\%$ on Mississippi, and
cos-max never exceeds $0.09$. Setting all three penalties to zero
(the $\lambda_X{=}0$ row, matching the NMF-FFB$_{\mathrm{unreg}}$
row of Table~\ref{tab:real_comp}) simultaneously raises the
off-ratio by one to two orders of magnitude and saturates
$\rho(X\Theta_1)$ at unity. Across the working range
$\lambda_X\in\{1,10,100\}$ the equilibrium MAE is identical to three
decimal places within each dataset, so approximate orthogonality is
achieved without a predictive-fit cost.

\begin{table}[htbp]
\centering
\small
\caption{Orthogonality verification on real data.
``off-ratio'' is the Frobenius-squared share of off-diagonal mass in
$X^{\top}X$; ``cos-max'' is the largest absolute column-cosine among
the latent basis columns.  Smaller is closer to orthogonal.
$\lambda_X{=}100$ is the value used throughout
Sections~\ref{subsec:simulation}--\ref{sec5}.
For each row, the basis $X$ is initialized via NNDSVDar followed by a
feedforward NMF-with-covariates fit at the same $\lambda_X$ used in
the final NMF-FFB fit, so that each row reflects a
self-consistent protocol from initialization to convergence.
The row $\lambda_X{=}0$ corresponds to the
NMF-FFB$_{\mathrm{unreg}}$ regime of Table~\ref{tab:real_comp},
in which all three penalties are simultaneously zero
($\lambda_X{=}\lambda_1{=}\lambda_2{=}0$); the resulting
$(\rho,\mathrm{MAE})$ values match Table~\ref{tab:real_comp} exactly,
and the row exhibits the expected spectral instability
$\rho(X\Theta_1)\!\approx\!1$.
The other rows ($\lambda_X\in\{1,10,100\}$) hold
$(\lambda_1,\lambda_2)$ at their cross-validated values from
Section~\ref{sec5} while varying only $\lambda_X$; small differences
from Table~\ref{tab:real_comp} in the last decimal reflect this
self-consistent orthogonality-scan protocol and rounding.}
\label{tab:ortho_real}
\begin{tabular}{llccccc}
\hline
Dataset & $\lambda_X$ & off-ratio & cos-max &
$\rho(X\Theta_1)$ & MAE \\
\hline
HS39
 & 0    & $4.4\times 10^{-2}$ & 0.180 & 1.002 & 2.971 \\
 & 1    & $1.2\times 10^{-2}$ & 0.118 & 0.361 & 0.181 \\
 & 10   & $1.1\times 10^{-2}$ & 0.110 & 0.360 & 0.181 \\
 & 100  & $6.1\times 10^{-3}$ & 0.079 & 0.357 & 0.181 \\
\hline
LA
 & 0    & $3.9\times 10^{-3}$ & 0.067 & 1.000 & 3.764 \\
 & 1    & $1.4\times 10^{-4}$ & 0.013 & 0.002 & 0.095 \\
 & 10   & $1.1\times 10^{-5}$ & 0.004 & 0.002 & 0.095 \\
 & 100  & $1.4\times 10^{-7}$ & 0.0004& 0.002 & 0.095 \\
\hline
Mississippi
 & 0    & $7.6\times 10^{-2}$ & 0.282 & 1.003 & 1.718 \\
 & 1    & $3.2\times 10^{-3}$ & 0.063 & 0.435 & 0.118 \\
 & 10   & $3.1\times 10^{-3}$ & 0.086 & 0.439 & 0.118 \\
 & 100  & $6.3\times 10^{-4}$ & 0.051 & 0.446 & 0.118 \\
\hline
\end{tabular}
\end{table}

Table~\ref{tab:ortho_sim} reports the same $\lambda_X$ scan on the
simulation grid of Section~\ref{subsec:simulation} under the
$N$-scaled penalty convention of \S\ref{subsubsec:Nscale} so that
the eight $(\rho_{\mathrm{true}}, N, \sigma)$ conditions can be
pooled at comparable effective regularization strength;
$R{=}500$ replications per condition.
The pattern mirrors the real-data result: at $\lambda_X{=}100$ the
mean off-ratio is $0.065\%$ and cos-max is below $0.03$, while
the penalty-free case $\lambda_X{=}0$ simultaneously raises the
off-ratio and saturates $\bar{\hat\rho}=0.916$. The
convergence of both scans on the same conclusion---adequate
orthogonality at $\lambda_X{=}100$ with no detectable predictive
cost---justifies fixing $\lambda_X{=}100$ as a working default for
the $[0,1]$-rescaled datasets in \S\ref{sec5} rather than retuning
it per dataset; the rationale and the adjustment rule for other
settings are stated in \S\ref{subsec:crossvalidation}.

\begin{table}[htbp]
\centering
\small
\caption{Orthogonality verification on the simulation grid of
Section~\ref{subsec:simulation} under the $N$-scaled penalty
convention (base $\lambda_X$ at $N_0{=}50$; effective penalty
scaled by $N/N_0$). HS39 dimensions $P_1{=}9$, $P_2{=}3$, $Q{=}3$;
$\sigma\in\{0,0.10\}$, $\rho_{\mathrm{true}}\in\{0,0.2\}$,
$N\in\{50,200\}$, eight conditions $\times$ $R{=}500$ replications,
pooled. The off-ratio and cos-max are defined as in
Table~\ref{tab:ortho_real}; $\bar{\hat\rho}$ is the mean estimated
spectral radius and $\overline{\mathrm{MAE}}$ the mean equilibrium
MAE.}
\label{tab:ortho_sim}
\begin{tabular}{cccccc}
\hline
$\lambda_X$ & off-ratio & cos-max & $\bar{\hat\rho}$ & $\overline{\mathrm{MAE}}$ \\
\hline
0    & $2.2\times 10^{-2}$ & 0.152 & 0.916 & 0.163 \\
1    & $2.0\times 10^{-2}$ & 0.141 & 0.032 & 0.041 \\
10   & $8.3\times 10^{-3}$ & 0.097 & 0.032 & 0.042 \\
100  & $6.5\times 10^{-4}$ & 0.027 & 0.037 & 0.043 \\
\hline
\end{tabular}
\end{table}

\section{Applications}\label{sec5}

\defcitealias{CHR2025}{UWPHI 2025}%

In this section we apply NMF-FFB to three real datasets to
demonstrate its empirical behavior and to illustrate how its
diagnostics---the spectral radius $\rho(X\Theta_1)$, the
amplification ratio (AR), and the structural correlations
$\mathrm{SC}_{\mathrm{map}}$ and $\mathrm{SC}_{\mathrm{cov}}$
introduced in Section~\ref{sec4}---translate into substantive
interpretations across different domains.
The three datasets (HS39, LA, Mississippi; abbreviations introduced
in \S\ref{sec1}) are summarized in Table~\ref{tab:datasets} and
treated in \S\ref{subsec:hs39}, \S\ref{subsec:pollution}, and
\S\ref{subsec:mshealth} respectively.
Together these applications cover a broad range of sample sizes
($N$ from $35$ to $508$) and feedback regimes (from near-zero
$\rho(X\Theta_1)$ on LA to moderate $\rho(X\Theta_1)\approx 0.45$
on Mississippi).
For each application we report the cross-validated hyperparameters,
the estimated basis matrix $X$ and structural matrices
$\Theta_1, \Theta_2$, and a path diagram of the resulting equilibrium
input--output mapping, with the qualitative interpretation discussed
in the corresponding subsection.

\paragraph*{Edge convention in path diagrams (Figs.\
\ref{fig:hs39_path}, \ref{fig:pollution_path},
\ref{fig:mshealth_path}).}
Each path diagram visualizes the three estimated matrices as three
distinct edge classes: arrows from latent factors to endogenous
outcomes $Y_1$ correspond to entries of the basis $X$
(forward read-out, column-stochastic); arrows from endogenous outcomes
$Y_1$ back to latent factors correspond to entries of the latent
feedback $\Theta_1$ (drawn only when above the display threshold $0.01$);
and arrows from exogenous drivers $Y_2$ to latent factors correspond
to entries of $\Theta_2$. A factor with no incoming $\Theta_1$ arrow
is therefore feedforward-only at the displayed precision.

The applications below use the \emph{fixed-penalty} convention
(per-dataset CV-selected $(\lambda_1,\lambda_2)$ with
$\lambda_X{=}100$); accordingly, absolute magnitudes of
$\hat\rho$ and $\widehat{\mathrm{AR}}$ across datasets at different
$N$ should be read using the $N$-scaled convention of
\S\ref{subsubsec:Nscale}, while the qualitative regime of each system
(near-zero feedback on LA, non-trivial feedback on HS39 and
Mississippi) is robust to this choice.

\begin{table}[htbp]
\centering
\caption{Summary of the three real datasets analyzed in
Section~\ref{sec5} (\S\ref{subsec:hs39}, \S\ref{subsec:pollution},
\S\ref{subsec:mshealth}).  $N$: sample size; $P_1$: number of
endogenous (outcome) indicators; $P_2$: number of exogenous
(covariate) indicators; $Q$: number of latent factors selected by
cross-validation.}
\label{tab:datasets}
\begin{tabular*}{\textwidth}{@{\extracolsep{\fill}}lccccl@{}}
\hline
Dataset & $N$ & $P_1$ & $P_2$ & $Q$ & Source \\
\hline
HS39         & 300 & 9  & 3  & 3 & \citet{holzinger1939} \\
LA           & 508 & 3  & 9  & 2 & astsa::lap \citep{astsa} \\
Mississippi  &  35 & 11 & 11 & 3 & \citepalias{CHR2025} \\
\hline
\end{tabular*}
\end{table}

\paragraph*{Stability of cross-validated hyperparameters.}
Hyperparameters $(\lambda_1, \lambda_2)$ are selected by 5-fold
equilibrium-MAE cross-validation (\S\ref{subsec:crossvalidation}) over
the grid $\lambda_k \in \{0,\,0.1,\,0.2,\,\ldots,\,0.9,\,1,\,2,\,\ldots,\,10\}$
(20 values, $400$ pairwise combinations).
To verify that the selection is not an artifact of grid resolution or
of an isolated local minimum of the CV surface, Table~\ref{tab:cv_stability}
reports the top three combinations for each application.
In all three applications the CV-selected pair lies on a flat region of
the CV surface: the top-three CV MAE values differ by less than
$0.05\%$ from the optimum, and adjacent combinations cluster tightly in
$(\lambda_1,\lambda_2)$ space (HS39 and LA select $\lambda_1=10$ across
the top three; Mississippi selects $\lambda_2=0.2$ across the top three).
The chosen $(\lambda_1,\lambda_2)$ is therefore robust to small grid
perturbations.

\begin{table}[htbp]
\centering
\small
\caption{Stability of cross-validated $(\lambda_1, \lambda_2)$ selection.
For each dataset, the top three combinations ordered by 5-fold
equilibrium-MAE cross-validation are shown.
The dagger ($\dagger$) marks the combination used in
Sections~\ref{subsec:hs39}--\ref{subsec:mshealth}.
$\Delta$\,CV\,MAE is the difference from the best value, in absolute
units and as a percentage of the best value.
(Here ``best/2nd/3rd'' refer to the CV-MAE ordering of penalty
combinations and are unrelated to the latent rank $Q$.)
The CV grid is
$\lambda_k \in \{0,\,0.1,\,0.2,\,\ldots,\,0.9,\,1,\,2,\,\ldots,\,10\}$
(20 values, 400 combinations).}
\label{tab:cv_stability}
\begin{tabular}{llccrr}
\hline
Dataset & Order & $\lambda_1$ & $\lambda_2$ & CV MAE &
$\Delta$ ($\%$) \\
\hline
HS39
 & Best ($\dagger$) & 10  & 0.6 & 0.18228 & ---\hspace{3pt}    \\
 & 2nd              & 10  & 0.7 & 0.18229 & 0.004\%  \\
 & 3rd              & 10  & 0.8 & 0.18229 & 0.005\%  \\
\hline
LA
 & Best ($\dagger$) & 10  & 0.8 & 0.09466 & ---\hspace{3pt}    \\
 & 2nd              & 10  & 0.6 & 0.09466 & 0.002\%  \\
 & 3rd              & 10  & 0.5 & 0.09467 & 0.010\%  \\
\hline
Mississippi
 & Best ($\dagger$) & 0.8 & 0.2 & 0.12721 & ---\hspace{3pt}    \\
 & 2nd              & 0.7 & 0.2 & 0.12725 & 0.031\%  \\
 & 3rd              & 0.9 & 0.2 & 0.12725 & 0.033\%  \\
\hline
\end{tabular}
\end{table}

Beyond the CV grid, increasing $\lambda_1$ further drives
$\sum\Theta_1$ to numerical zero (already at $\lambda_1{=}10$ for the
small-$N$ Mississippi application and at $\lambda_1{\ge}20$ for HS39
and LA), at which point NMF-FFB collapses onto the NMF-FF column of
Table~\ref{tab:real_comp}; the chosen grid upper bound thus brackets
the transition between non-trivial feedback and full feedforward
collapse.

Throughout this section the path diagrams summarize the equilibrium
input--output mapping $Y_2\mapsto Y_1$ rather than a
covariance-based system: arrows represent non-negative additive
contributions, with latent-factor-to-$Y_1$ edges encoding the basis
matrix $X$ (column-stochastic), $Y_1$-to-latent-factor edges
encoding the latent feedback coefficients $\Theta_1$, and
$Y_2$-to-latent-factor edges encoding the exogenous drivers
$\Theta_2$.

\paragraph*{Status of zeros in the estimated matrices.}
In NMF-FFB the entries of $\Theta_1$ and $\Theta_2$ are treated as
\emph{free non-negative parameters}, and the entries of $X$ are
treated as free non-negative parameters subject to the
column-stochastic constraint $\mathbf{1}^\top X = \mathbf{1}^\top$
introduced in \S\ref{subsec:basic}; unlike Covariance-Based SEM
we do not impose any a-priori zero-path constraints analogous to
fixed loadings. Accordingly, values shown as $0.00$ in the basis, $\Theta_2^\top$,
and structural-coefficient tables that follow (i.e.,
Tables~\ref{tab:hs39_basis}--\ref{tab:ms_struct_compare})
should be read as estimates lying at or very near the non-negative
boundary after rounding, rather than parameters fixed to zero by the
model specification. In regularized columns, the mechanism producing each
zero differs by matrix: in $\Theta_1$ and $\Theta_2$, zeros arise
primarily from $\ell_1$-induced shrinkage combined with the
non-negativity constraint; in $X$, $\ell_1$ is not applied, and
zeros arise from the joint action of the non-negativity constraint,
the orthogonality penalty $\tfrac{\lambda_X}{2}\|X^\top X -
\mathrm{diag}(X^\top X)\|_F^2$ which separates basis columns into
near-disjoint supports, the column-stochastic normalization, and
two-decimal display rounding. In unregularized columns, shown only for
the HS39 diagnostic comparison, $0.00$ denotes numerical zeros after
rounding or degeneracy of the penalty-free solution rather than
penalty-induced shrinkage.

Table~\ref{tab:summary_applications} summarizes the metrics defined
in Section~\ref{sec4}, with $95\%$ bootstrap percentile intervals
for $\rho(X\Theta_1)$ and AR. The three datasets together cover
moderate latent amplification (HS39), near-feedforward behavior (LA),
and moderate $\rho(X\Theta_1)$ but small cumulative amplification
(Mississippi), and absolute-magnitude
statements are reserved for the within-dataset discussion of
\S\ref{subsec:hs39}--\S\ref{subsec:mshealth}.

\begin{table}[htbp]
\centering
\small
\caption{
Structural and predictive metrics for the three empirical datasets.
Values in brackets denote 95\% X-fixed bootstrap percentile intervals
based on resampling the $(Y_1,Y_2)$ columns and re-estimating
$(\Theta_1,\Theta_2)$ conditional on the original-data $X$ and fixed
hyperparameters. Sample sizes were
$N=300$ (HS39), $N=508$ (LA), and $N=35$ (Mississippi), with latent
dimensions $Q=3,2,$ and $3$, respectively.
}
\label{tab:summary_applications}
\begin{tabular*}{\textwidth}{@{\extracolsep{\fill}}lccccc@{}}
\hline
Dataset & $\rho(X\Theta_1)$ & AR & SC$_{\mathrm{map}}$ &
SC$_{\mathrm{cov}}$ & MAE \\
\hline
HS39
& 0.357 $[0.25,\,0.46]$
& 1.361 $[1.23,\,1.49]$
& 0.999 & 0.980 & 0.181 \\
LA
& 0.002 $[0.00,\,0.01]$
& 1.002 $[1.00,\,1.01]$
& 0.882 & 0.983 & 0.095 \\
Mississippi
& 0.446 $[0.09,\,0.56]$
& 1.025 $[1.00,\,1.09]$
& 0.950 & 0.989 & 0.117 \\
\hline
\end{tabular*}
\end{table}

Bootstrap percentile intervals confirm that the estimates of 
$\rho(X\Theta_1)$ and AR are reasonably concentrated for HS39 and LA,
whereas the Mississippi intervals remain wider and should be interpreted
more cautiously.

The remainder of this section examines each dataset in turn, focusing on
the recovered latent components, the estimated exogenous pathways, and
the role of feedback in shaping cumulative effects.

\subsection{Holzinger--Swineford cognitive-ability analysis}
\label{subsec:hs39}

The first application uses the classical Holzinger--Swineford (HS39)
cognitive-test dataset \citep{holzinger1939}, a standard benchmark in
factor analysis \citep{joreskog1969}.  
Nine test scores were treated as endogenous variables, and three
background variables (reversed age, gender, school indicator) were used
as exogenous inputs after rescaling all variables to $[0,1]$.
Here ``reversed age'' refers to the variable $\max(\text{age}) -
\text{age}$, applied so that larger values of the exogenous input
correspond to \emph{younger} pupils and the non-negativity convention
``larger input is a larger positive driver of the outcome'' is preserved
for all exogenous variables.
NMF-FFB was fitted with latent dimension $Q=3$, corresponding to the
established three-factor structure.

The estimated basis matrix $X$ (Table~\ref{tab:hs39_basis})
recovers the canonical grouping of visual (\texttt{x1--x3}), verbal
(\texttt{x4--x6}), and speeded-performance (\texttt{x7--x9}) tests,
showing that NMF-FFB extracts the classical ability pattern in a
non-negative and data-driven manner.
To make the role of regularization concrete we display, alongside
the CV-optimal estimates, the corresponding matrices fitted with
\emph{all three penalties set to zero}
($\lambda_X{=}\lambda_1{=}\lambda_2{=}0$); the latter is the same
NMF-FFB$_{\mathrm{unreg}}$ regime as the second row of HS39 in
Table~\ref{tab:real_comp}.

\begin{table}[htbp]
  \centering
  \caption{Estimated basis matrix $X$ for the HS39 data ($Q=3$).
  CV-optimal columns use $(\lambda_X,\lambda_1,\lambda_2){=}(100,10,0.6)$;
  unregularized columns use $\lambda_X{=}\lambda_1{=}\lambda_2{=}0$.
  Variables follow \texttt{lavaan::HolzingerSwineford1939}:
  \texttt{x1} visual perception,
  \texttt{x2} cubes,
  \texttt{x3} lozenges,
  \texttt{x4} paragraph comprehension,
  \texttt{x5} sentence completion,
  \texttt{x6} word meaning,
  \texttt{x7} speeded addition,
  \texttt{x8} speeded counting of dots,
  \texttt{x9} speeded discrimination of straight and curved
  capitals. Values shown as $0.00$ are estimates at or near the
  non-negative boundary after rounding, not parameters fixed to zero
  (see \S\ref{sec5} preamble); in the unregularized columns they
  denote numerical zeros after rounding rather than penalty-induced
  shrinkage.}
  \label{tab:hs39_basis}
  \begin{tabular}{l|ccc|ccc}
    \hline
     & \multicolumn{3}{c|}{CV-optimal}
     & \multicolumn{3}{c}{Unregularized ($\lambda{=}0$)} \\
    Variable & Factor 1 & Factor 2 & Factor 3 & Factor 1 & Factor 2 & Factor 3 \\
    \hline
    \texttt{x1} (visual)        & 0.27 & 0.03 & 0.00 & 0.17 & 0.11 & 0.12 \\
    \texttt{x2} (cubes)         & 0.27 & 0.02 & 0.00 & 0.25 & 0.10 & 0.08 \\
    \texttt{x3} (lozenges)      & 0.26 & 0.00 & 0.00 & 0.52 & 0.00 & 0.00 \\
    \texttt{x4} (paragraph)     & 0.00 & 0.31 & 0.00 & 0.00 & 0.26 & 0.00 \\
    \texttt{x5} (sentence)      & 0.00 & 0.36 & 0.00 & 0.00 & 0.31 & 0.00 \\
    \texttt{x6} (word)          & 0.00 & 0.22 & 0.00 & 0.00 & 0.20 & 0.00 \\
    \texttt{x7} (addition)      & 0.00 & 0.00 & 0.65 & 0.00 & 0.00 & 0.33 \\
    \texttt{x8} (counting dots) & 0.05 & 0.03 & 0.28 & 0.00 & 0.00 & 0.25 \\
    \texttt{x9} (capitals)      & 0.15 & 0.04 & 0.07 & 0.06 & 0.01 & 0.22 \\
    \hline
  \end{tabular}
\end{table}

\begin{table}[htbp]
  \centering\small
  \caption{Exogenous driver coefficients $\Theta_2^\top$ for HS39
  (rows: exogenous variables; columns: latent factors).
  CV-optimal vs.\ unregularized ($\lambda{=}0$) regimes.
  In the CV-optimal columns, values shown as $0.00$ are penalized
  estimates at or near the non-negative boundary; in the unregularized
  columns, they denote numerical zeros after rounding or the
  degeneracy of the penalty-free solution, not penalty-induced
  shrinkage (see \S\ref{sec5} preamble).
  Bootstrap-based one-sided significance markers for the CV-optimal
  entries are reported instead in
  Table~\ref{tab:hs39_struct_compare}.}
  \label{tab:hs39_theta2}
  \begin{tabular}{l|ccc|ccc}
    \hline
     & \multicolumn{3}{c|}{CV-optimal}
     & \multicolumn{3}{c}{Unregularized ($\lambda{=}0$)} \\
    Variable & Factor 1 & Factor 2 & Factor 3 & Factor 1 & Factor 2 & Factor 3 \\
    \hline
    \texttt{age.rev}   & 2.01 & 1.56 & 0.60 & 0.00 & 0.00 & 0.00 \\
    \texttt{sex.2}     & 0.06 & 0.15 & 0.11 & 0.00 & 0.01 & 0.01 \\
    \texttt{school.GW} & 0.04 & 0.24 & 0.00 & 0.00 & 0.05 & 0.00 \\
    \hline
  \end{tabular}
\end{table}

The CV/unregularized contrast in
Tables~\ref{tab:hs39_basis} and~\ref{tab:hs39_theta2} is striking:
without regularization, $X$ loses its column-disjoint block
structure, the latent feedback operator $\Theta_1$ becomes dense
with several entries well above unity (e.g., the largest entries are
$1.44, 1.36, 1.34, 1.10, 1.06$ in the unregularized regime, vs.\
$\le 0.95$ under CV), and $\Theta_2$ collapses to numerical zero so
that the exogenous channel switches off. Combined with
$\rho(X\Theta_1)\!\approx\!1.00$ (Table~\ref{tab:real_comp}), the
model degenerates into the trivial fixed point
$Y_1\!\approx\!X\Theta_1 Y_1$ with MAE $2.97$ instead of $0.18$. The
CV-selected penalties are therefore essential rather than cosmetic:
they keep the exogenous channel alive and prevent the collapse onto
this trivial high-feedback solution.

\begin{figure}[htbp]
  \centering
  \includegraphics[width=0.98\linewidth]{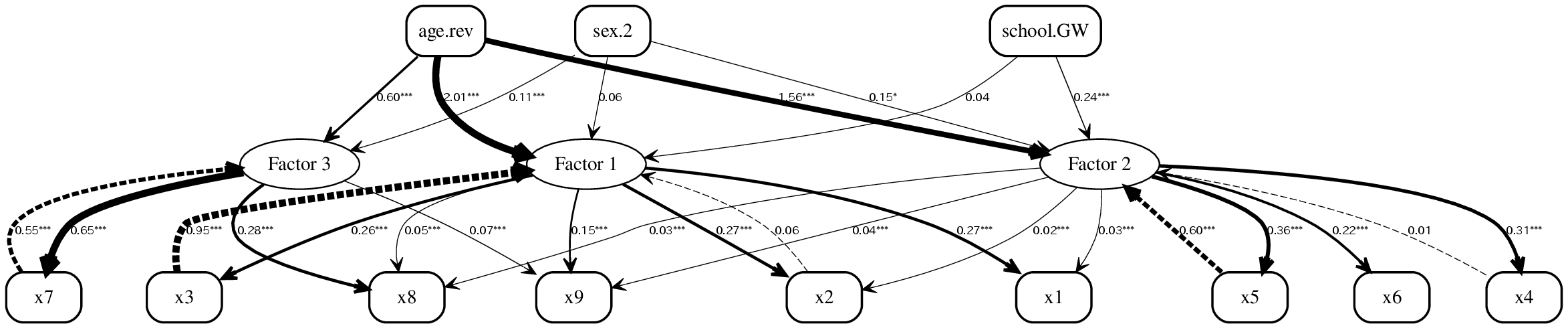}
  \caption{
    Path diagram estimated by NMF-FFB ($Q=3$) for HS39.
    Endogenous variables $Y_1$: cognitive tests
    \texttt{x1}--\texttt{x9}; three latent factors; exogenous
    variables $Y_2$: \texttt{age.rev} (reversed age),
    \texttt{sex.2} (sex), \texttt{school.GW} (school).
    Edge convention: factor $\to Y_1$ edges encode $X$,
    $Y_1 \to$ factor edges encode $\Theta_1$ (latent feedback),
    and $Y_2 \to$ factor edges encode $\Theta_2$
    (exogenous channel).
    The system exhibits moderate latent feedback (amplification
    ratio $\approx 1.36$). Under the CV-selected penalties, the
    only non-trivial latent-feedback entries are the loadings of
    the lozenges (\texttt{x3}), sentence (\texttt{x5}), and
    addition (\texttt{x7}) tests onto the visual, verbal, and speed
    factors, respectively (values $0.95$, $0.60$, $0.55$); all
    other $\Theta_1$ entries are negligible at the $0.01$ display
    threshold.
    Significance markers ($*$, $**$, $***$) on edge labels are
    one-sided support-rate tests at thresholds $0.95$, $0.99$,
    $0.999$ on $B_{\mathrm{boot}}{=}1000$ $X$-fixed bootstrap replicates
    (see \S\ref{subsec:feedback_metrics}).}
  \label{fig:hs39_path}
\end{figure}

The estimated path diagram is shown in
Figure~\ref{fig:hs39_path}.
The latent-feedback operator $X\Theta_1$ was sparse, with feedback pathways from representative variables to their associated factors (e.g.,
\texttt{x3} for visual, \texttt{x5} for verbal, \texttt{x7} for speed).
Feedback strength was moderate in spectral-radius terms, with
$\rho(X\Theta_1)\approx 0.35$ and amplification ratio
$\mathrm{AR}\approx 1.36$. We caveat this reading explicitly:
Table~\ref{tab:real_comp} shows that the equilibrium MAE of NMF-FFB
on HS39 is essentially indistinguishable from that of the
feedforward ablation NMF-FF with $\Theta_1=0$ (both $0.181$), so the
case for non-zero latent feedback in HS39 does not rest on
predictive improvement over the feedforward variant. Rather, it
rests on (i) the bootstrap percentile interval for $\rho(X\Theta_1)$
excluding zero (Table~\ref{tab:summary_applications}), and (ii) the
sparse, interpretable feedback pattern in $X\Theta_1$ noted above.
The HS39 evidence for latent feedback is therefore best read as
combinatorial---bootstrap support and interpretable sparsity, rather
than predictive gain.

Exogenous inputs showed expected patterns, confirmed at the
bootstrap one-sided level (Table~\ref{tab:hs39_struct_compare}):
reversed age affects all three factors at $p{<}0.001$
($\Theta_2$ on visual, verbal, speed: $2.01, 1.56, 0.60$), school
membership exhibits a verbal-specific pathway at $p{<}0.001$
($\Theta_2 = 0.24$ on verbal; n.s.\ on visual and speed), and gender
differences concentrate on the speeded-performance factor at
$p{<}0.01$ ($\Theta_2 = 0.11$ on speed) with a smaller verbal
contribution at $p{<}0.05$ ($0.15$).
The latent-feedback entries are also bootstrap-significant: the
self-reinforcing loadings $\texttt{x3}\!\to\!\text{Factor 1}$
(support rate $1.000$, $p{<}0.001$),
$\texttt{x5}\!\to\!\text{Factor 2}$ (support rate $0.999$,
$p{<}0.01$), and $\texttt{x7}\!\to\!\text{Factor 3}$
(support rate $0.999$, $p{<}0.01$) all reach bootstrap support
exceeding $0.99$ ($\Theta_1$ values $0.95, 0.60, 0.55$ in
Fig.~\ref{fig:hs39_path}).
Structural fidelity was high: IO structural correlation exceeded $0.99$,
covariance structural correlation was near $0.98$, and the equilibrium
MAE was approximately $0.18$.
This MAE is essentially tied with the saturated signed OLS-MIMIC
benchmark on the same data (Table~\ref{tab:real_comp},
MAE${}={}0.180$), demonstrating that the parts-based non-negative
decomposition of NMF-FFB achieves predictive accuracy comparable to
an unconstrained linear mapping with $P_1 P_2$ free parameters, while
additionally delivering an interpretable equilibrium structure that
OLS-MIMIC does not provide.

Overall, the HS39 results demonstrate that NMF-FFB recovers the familiar
three-factor structure while quantifying modest latent feedback in a
non-negative framework.

\paragraph*{Reading the latent Leontief inverse.}
For HS39, the Leontief inverse $(I-X\hat\Theta_1)^{-1}$ implies that
the direct exogenous inputs to the three ability-related factors are
moderately amplified through repeated latent feedback. Because
$\hat\rho(X\hat\Theta_1)\approx 0.35$, the Neumann expansion converges
rapidly, but the higher-order terms are not negligible: they raise the
total effect by about $36\%$ relative to the direct component
$X\hat\Theta_2$. In substantive terms, this means that age- and
school-related inputs are not simply mapped once into latent abilities;
their effects are recycled through the latent-factor structure before
equilibrium is reached.

\emph{Numerical summary of $(I - X\hat\Theta_1)^{-1}$.}
The $9\times 9$ matrix has diagonal entries in the range $1.00$ to
$1.45$ (mean $1.10$) and absolute off-diagonal entries in the range
$0$ to $0.39$ (mean $0.02$), so self-amplification on the diagonal
dominates and cross-variable propagation is concentrated on a few
entries consistent with the three-factor block structure.
The Frobenius norms of the first four Neumann-expansion terms
$\{I,\, X\hat\Theta_1,\, (X\hat\Theta_1)^2,\, (X\hat\Theta_1)^3\}$ are
$3.00, 0.58, 0.15, 0.045$, giving a rapid geometric decay
($\approx 1/(1-\hat\rho)=1.56$ as the total series-norm bound).
Thus the direct term $X\hat\Theta_2$ accounts for most of the
equilibrium mapping, with the first two feedback echoes providing the
bulk of the remaining $\approx 36\%$ correction.

\paragraph*{Comparison with standard SEM.}
For an external benchmark, we also fit the canonical three-factor
MIMIC model on HS39 with the \texttt{lavaan} package
\citep{lavaan}, specifying the same measurement structure
($\mathrm{Factor}_q$ measured by
$x_{3q-2},x_{3q-1},x_{3q}$ for $q{=}1,2,3$) and regressing each
factor on the same exogenous variables (\texttt{age.rev},
\texttt{sex.2}, \texttt{school.GW}).
The model converges with moderate fit
($\chi^2(42){=}149.1$,
$\mathrm{CFI}{=}0.893$, $\mathrm{TLI}{=}0.840$,
$\mathrm{RMSEA}{=}0.092$ [90\%\,CI $0.076$--$0.108$],
$\mathrm{SRMR}{=}0.066$), which places HS39 close to but below the
conventional ``acceptable fit'' thresholds of
$\mathrm{CFI}{\ge}0.90$ and $\mathrm{RMSEA}{\le}0.08$.
Table~\ref{tab:hs39_struct_compare} compares the structural
coefficients (factor $\leftarrow$ exogenous variable) between
SEM (signed, standardized) and NMF-FFB
($\Theta_2$, non-negative).

\begin{table}[htbp]
  \centering\small
  \caption{HS39 structural coefficients: standard SEM vs.\ NMF-FFB.
  Each factor block reports the SEM standardized regression coefficient
  $\hat\beta$ with two-sided significance markers
  (\textsuperscript{***}$p{<}0.001$,
  \textsuperscript{**}$p{<}0.01$,
  \textsuperscript{*}$p{<}0.05$,
  \textsuperscript{.}$p{<}0.1$, otherwise n.s.;
  $H_1{:}\beta\ne 0$) alongside the CV-optimal non-negative
  $\Theta_2$ entry from NMF-FFB with one-sided support-rate
  significance markers ($H_1{:}\theta_2{>}0$;
  $*$, $**$, $***$ denote bootstrap support
  rates $\Pr_{\mathrm{boot}}(\hat\theta_2^{(b)}{>}0.01)$ exceeding
  $0.95$, $0.99$, $0.999$ on $B_{\mathrm{boot}}{=}1000$ $X$-fixed
  bootstrap replicates; see \S\ref{subsec:feedback_metrics}).}
  \label{tab:hs39_struct_compare}
  \begin{tabular}{l@{\hskip 6pt}r@{\hskip 6pt}r@{\hskip 14pt}r@{\hskip 6pt}r@{\hskip 14pt}r@{\hskip 6pt}r}
    \hline
                  & \multicolumn{2}{c}{Factor 1} & \multicolumn{2}{c}{Factor 2} & \multicolumn{2}{c}{Factor 3} \\
                  & \multicolumn{2}{c}{(visual)} & \multicolumn{2}{c}{(verbal)} & \multicolumn{2}{c}{(speed)}  \\
    Variable      & \multicolumn{1}{c}{SEM} & \multicolumn{1}{c}{NMF-FFB} & \multicolumn{1}{c}{SEM} & \multicolumn{1}{c}{NMF-FFB} & \multicolumn{1}{c}{SEM} & \multicolumn{1}{c}{NMF-FFB} \\
    \hline
    \texttt{age.rev}    & $\phantom{-}0.07$\phantom{\textsuperscript{***}}    & $2.01$\textsuperscript{***}              & $\phantom{-}0.18$\textsuperscript{**}\phantom{*}    & $1.56$\textsuperscript{***}              & $-0.26$\textsuperscript{***}                        & $0.60$\textsuperscript{***}              \\
    \texttt{sex.2}      & $-0.21$\textsuperscript{**}\phantom{*}              & $0.06$\phantom{\textsuperscript{***}}    & $\phantom{-}0.06$\phantom{\textsuperscript{***}}    & $0.15$\textsuperscript{*}\phantom{**}    & $\phantom{-}0.08$\phantom{\textsuperscript{***}}    & $0.11$\textsuperscript{**}\phantom{*}    \\
    \texttt{school.GW}  & $-0.13$\phantom{\textsuperscript{***}}              & $0.04$\phantom{\textsuperscript{***}}    & $\phantom{-}0.24$\textsuperscript{***}              & $0.24$\textsuperscript{***}              & $-0.07$\phantom{\textsuperscript{***}}              & $0.00$\phantom{\textsuperscript{***}} \\
    \hline
  \end{tabular}
\end{table}

\textit{Data-driven recovery of the canonical three-factor structure.}
The principal methodological point of HS39 lies \emph{upstream} of
the structural-coefficient table. Standard SEM \emph{requires} the
analyst to pre-specify which observed variables load on which latent
factor---here,
$\mathrm{Factor}_q = \{x_{3q-2}, x_{3q-1}, x_{3q}\}$ for $q=1,2,3$%
---a confirmatory setup that depends on a century of accumulated
knowledge of the Holzinger--Swineford battery. NMF-FFB takes only
the latent dimension $Q=3$ as input and \emph{discovers} the
assignment of observed variables to latent factors data-drivenly,
through the basis matrix $X$ (Table~\ref{tab:hs39_basis}). The
recovered $X$ largely recapitulates the canonical three-factor
structure: $\{x_1,x_2,x_3\}$ load most strongly on Factor~1,
$\{x_4,x_5,x_6\}$ on Factor~2, and the speed block is concentrated
on Factor~3 for \texttt{x7} and \texttt{x8}, while \texttt{x9} is
split between Factor~1 and Factor~3 in the CV-optimal solution, with
no measurement structure supplied to the estimator. HS39 therefore
serves as evidence that the proposed exploratory NMF-FFB recovers the
main block structure that standard SEM can fit only after the
structure has been hand-specified, while retaining the small ambiguity
visible for \texttt{x9}. This upstream contribution
propagates to the next two subsections: the SEM measurement-model
groupings used as benchmarks for LA (\S\ref{subsec:pollution}) and
Mississippi (\S\ref{subsec:mshealth}) are themselves borrowed from
the basis matrix $X$ that NMF-FFB discovers.

\textit{Coefficient-level remarks.}
At the structural-coefficient level the two methods agree most
strongly where the underlying signed effect is positive and large
(\texttt{school.GW} $\to$ verbal factor: SEM $+0.24$ at $p{<}0.001$
vs.\ NMF-FFB $\Theta_2=0.24$, identical to two decimals). Where SEM
identifies sign-bearing effects (\texttt{age.rev} $\to$ speed
$=-0.26$, $p{<}0.001$; \texttt{sex.2} $\to$ visual
$=-0.21$, $p{<}0.01$), NMF-FFB cannot represent the sign
and either folds the magnitude into a non-negative load on every
factor (\texttt{age.rev} appears as the dominant single driver of
all three factors at $2.01, 1.56, 0.60$) or shrinks the entry near
zero (\texttt{sex.2} $\to$ visual $=0.06$). On HS39 the
methodological payoff of NMF-FFB is therefore not in the structural
coefficients but in the upstream data-driven measurement structure
described above, together with the additional outputs $\Theta_1$
(latent feedback) and the Leontief inverse $(I-X\Theta_1)^{-1}$
that standard SEM does not produce.

\subsection{LA pollution--mortality analysis}
\label{subsec:pollution}

We analyzed the \texttt{lap} dataset from the \texttt{astsa} package
\citep{astsa}, which contains weekly mortality counts together with
climatic variables and pollutant concentrations (CO, SO$_2$, NO$_2$,
hydrocarbons, ozone, particulates) from 1970--1979.
Endogenous variables ($Y_1$) consist of total, respiratory, and
cardiovascular mortality; exogenous variables ($Y_2$) include temperature,
humidity, pollutant levels, and a quadratic temperature term.
Mortality counts were log-transformed and all variables were rescaled to
$[0,1]$ with signs adjusted so that larger values indicate higher risk.

Cross-validation selected a latent dimension of $Q=2$.
NMF-FFB was fitted using $\lambda_X=100$ and sparsity parameters
$(\lambda_1,\lambda_2)$ chosen to minimize the 5-fold equilibrium MAE.
The estimated latent-feedback operator was extremely weak:
$\rho(X\Theta_1)\approx 0.002$, and the amplification ratio was nearly
one (AR$\approx 1.002$), indicating that weekly mortality responds almost
entirely to direct environmental inputs rather than latent feedback.

The estimated basis matrix $X$ (Table~\ref{tab:pollution_basis})
separates respiratory mortality (Factor~1) from the combined pattern of
cardiovascular and total mortality (Factor~2).
All entries of the latent feedback coefficients $\Theta_1^\top$
are zero to two decimal places under the CV-selected penalties,
consistent with the near-zero spectral radius
$\hat\rho(X\Theta_1)\!\approx\!0.002$; the entire fit is therefore
carried by the exogenous coefficients $\Theta_2^\top$, whose
CV-optimal values are reported alongside SEM in
Table~\ref{tab:la_struct_compare}.

\begin{table}[htbp]
  \centering
  \caption{Estimated basis matrix $X$ for LA pollution data ($Q=2$),
  CV-optimal $(\lambda_X,\lambda_1,\lambda_2){=}(100,10,0.8)$.
  Values shown as $0.00$ are penalized estimates at or near the
  non-negative boundary (see \S\ref{sec5} preamble).}
  \label{tab:pollution_basis}
  \begin{tabular}{lcc}
    \hline
    Variable & Factor 1 & Factor 2 \\
    \hline
    \texttt{rmort} (respiratory mortality)   & 1.00 & 0.00 \\
    \texttt{tmort} (total mortality)         & 0.00 & 0.47 \\
    \texttt{cmort} (cardiovascular mortality)& 0.00 & 0.53 \\
    \hline
  \end{tabular}
\end{table}

\begin{figure}[htbp]
  \centering
  \includegraphics[width=0.98\linewidth]{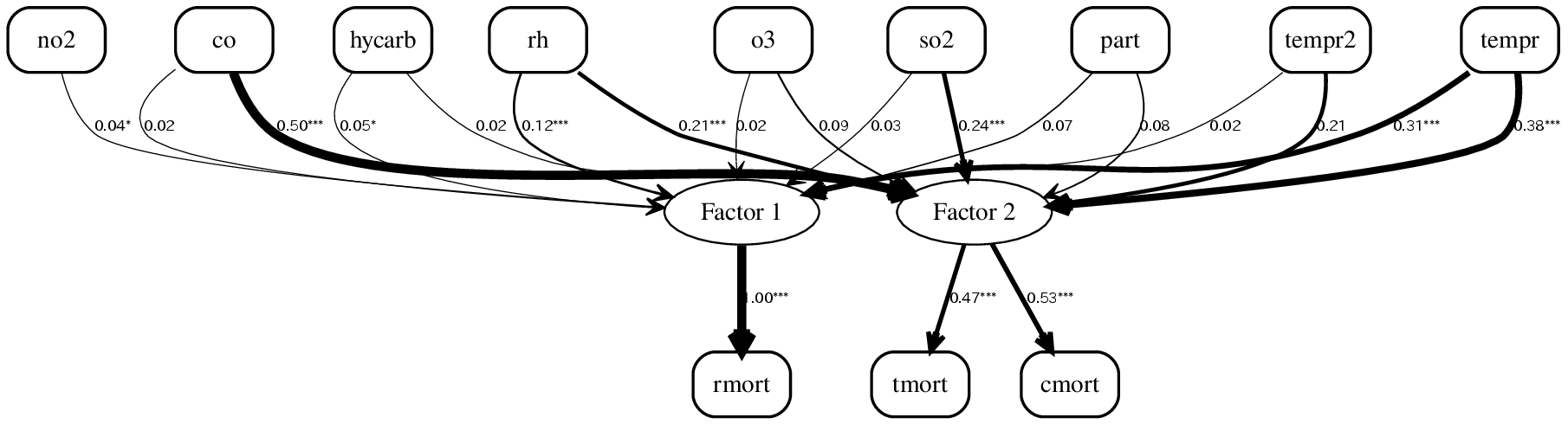}
  \caption{
    Path diagram estimated by NMF-FFB ($Q=2$) for the LA
    pollution--mortality system.
    Endogenous variables $Y_1$: \texttt{tmort} (total mortality),
    \texttt{rmort} (respiratory mortality), \texttt{cmort}
    (cardiovascular mortality); two latent factors; exogenous
    variables $Y_2$: \texttt{tempr} (temperature), \texttt{rh}
    (relative humidity), \texttt{co} (carbon monoxide),
    \texttt{so2} (sulfur dioxide), \texttt{no2} (nitrogen dioxide),
    \texttt{hycarb} (hydrocarbons), \texttt{o3} (ozone),
    \texttt{part} (particulates), and \texttt{tempr2} (squared
    temperature).
    Edge convention: factor $\to Y_1$ edges encode $X$,
    $Y_1 \to$ factor edges encode $\Theta_1$ (latent feedback),
    and $Y_2 \to$ factor edges encode $\Theta_2$
    (exogenous channel).
    Factor~1 reflects climatic drivers of respiratory mortality,
    whereas Factor~2 captures pollutant-related variation in
    cardiovascular and total mortality.
    Latent feedback is negligible
    ($\hat\rho(X\hat\Theta_1)\!\approx\!0.002$, all $\Theta_1$ entries
    below the $0.01$ display threshold), so no return arrows from
    factors to outcomes appear in the diagram.
    Significance markers ($*$, $**$, $***$) on edge labels are
    one-sided support-rate tests at thresholds $0.95$, $0.99$,
    $0.999$ on $B_{\mathrm{boot}}{=}1000$ $X$-fixed bootstrap replicates
    (see \S\ref{subsec:feedback_metrics}).}
  \label{fig:pollution_path}
\end{figure}

The structural diagram (Figure~\ref{fig:pollution_path}) shows that
Factor~1 is primarily activated by climatic variables, whereas Factor~2
combines the cardiovascular pollution channel with climatic covariates
that are strongly collinear with the weekly pollution series.
Inspection of the leading $\Theta_2$ entries makes this split concrete.
For Factor~1, the largest exogenous drivers are temperature and
relative humidity, followed by smaller pollutant entries, consistent
with a climate-modulated respiratory channel in which thermal stress
and humidity dominate weekly variation. For Factor~2, the largest
entries are CO, temperature, SO$_2$, relative humidity, and the
quadratic temperature term; particulates and ozone are smaller
positive entries, while NO$_2$ and hydrocarbons are weak after the
regularized non-negative selection. Thus $\Theta_2$ does not support
a clean pollutant-only interpretation of Factor~2; rather, it separates
a respiratory climate-dominated factor from a cardiovascular factor in
which combustion-related pollution proxies and collinear climate terms
jointly enter.

By contrast, refitting LA with all three penalties set to zero
($\lambda_X{=}\lambda_1{=}\lambda_2{=}0$; second LA row of
Table~\ref{tab:real_comp}) yields $\hat\rho(X\Theta_1)\!\approx\!1.00$
and the entire $\Theta_2$ collapses to numerical zero, so $X\Theta_1$
acquires an eigenvalue $\approx 1$ on the column space of $Y_1$ and
the model degenerates into the trivial fixed point
$Y_1 \approx X\Theta_1 Y_1$ (MAE $3.76$).

These pathways are corroborated by NMF-FFB's bootstrap inference at
$p{<}0.001$ (CO and SO$_2$ on the cardiovascular factor:
$\Theta_2 = 0.50$ and $0.24$, respectively;
Table~\ref{tab:la_struct_compare}) and align with epidemiological
evidence that pollutants elevate cardiovascular risk
\citep{brook2010}; NMF-FFB recovers this distinction in a fully
data-driven manner while confirming negligible latent feedback (all
$\Theta_1$ entries bootstrap-non-significant, consistent with
$\hat\rho(X\Theta_1)\!\approx\!0.002$).

Overall, the analysis indicates that NMF-FFB identifies two distinct
mortality components and that latent feedback is negligible in this
setting; the five-method comparison in
Table~\ref{tab:real_comp} confirms that the predictive fit of NMF-FFB
on LA matches that of NMF-FF (which sets $\Theta_1{=}0$ a priori)
and is essentially tied with both the saturated OLS-MIMIC baseline
and the Standard SEM benchmark (lavaan ML).

\paragraph*{Reading the latent Leontief inverse.}
Because $\hat\rho(X\hat\Theta_1)\approx 0.002$ on LA, the equilibrium
operator $(I-X\hat\Theta_1)^{-1}$ reduces to essentially the identity
and the higher-order terms of the Neumann expansion
$I + X\hat\Theta_1 + (X\hat\Theta_1)^{2} + \cdots$ are numerically
negligible; the total effect of a pollutant shock on mortality
therefore coincides with the direct effect $X\hat\Theta_2$, leaving
no room for Leontief-type amplification in this system.

\paragraph*{Comparison with standard SEM.}
For an external benchmark, we also fit a path-analytic SEM on LA with
the \texttt{lavaan} package \citep{lavaan}, mirroring the NMF-FFB
grouping in $X$ (Table~\ref{tab:pollution_basis}). NMF-FFB treats
respiratory mortality (\texttt{rmort}) as a one-element factor of
its own, separate from the two-element group of total mortality
(\texttt{tmort}) and cardiovascular mortality (\texttt{cmort}). The
standard SEM analogue of this structure has only one true latent
factor: a \emph{cardiovascular factor} that summarizes the joint
behavior of \texttt{tmort} and \texttt{cmort}, while \texttt{rmort}
is retained as a directly observed outcome. Both the cardiovascular
latent factor and the observed \texttt{rmort} are then regressed on
the same nine exogenous variables (temperature, humidity, the four
pollutants, ozone, particulates, and the quadratic temperature
term). This is a mixed observed--latent (``hybrid path-analytic'')
SEM specification rather than a pure two-factor MIMIC, but it is the
direct SEM counterpart of the NMF-FFB grouping and avoids any
artificial single-indicator latent factor.
The model converges but the fit is poor:
$\chi^2(9){=}179.8$,
$\mathrm{CFI}{=}0.908$, $\mathrm{TLI}{=}0.692$,
$\mathrm{RMSEA}{=}0.193$ [90\%\,CI $0.169$--$0.218$],
$\mathrm{SRMR}{=}0.018$, falling well outside conventional acceptable
ranges and reflecting strong multicollinearity among the regressors
(see below). Table~\ref{tab:la_struct_compare} compares the
structural coefficients (outcome $\leftarrow$ exogenous variable)
between SEM (signed, standardized) and NMF-FFB
($\Theta_2$, non-negative).

\begin{table}[htbp]
  \centering\small
  \caption{LA structural coefficients: standard SEM (hybrid
  path-analytic specification: \texttt{rmort} as observed outcome,
  cardiovascular factor as the latent factor of \{\texttt{tmort},
  \texttt{cmort}\}) vs.\ NMF-FFB. SEM coefficients are standardized;
  significance markers as in Table~\ref{tab:hs39_struct_compare}.}
  \label{tab:la_struct_compare}
  \begin{tabular}{l@{\hskip 6pt}r@{\hskip 6pt}r@{\hskip 14pt}r@{\hskip 6pt}r}
    \hline
                  & \multicolumn{2}{c}{Factor 1}      & \multicolumn{2}{c}{Factor 2} \\
                  & \multicolumn{2}{c}{(respiratory)} & \multicolumn{2}{c}{(cardiovascular)} \\
    Variable      & \multicolumn{1}{c}{SEM} & \multicolumn{1}{c}{NMF-FFB} & \multicolumn{1}{c}{SEM} & \multicolumn{1}{c}{NMF-FFB} \\
    \hline
    \texttt{tempr}   & $\phantom{-}2.32$\textsuperscript{***}              & $0.31$\textsuperscript{***}              & $\phantom{-}3.81$\textsuperscript{***}              & $0.38$\textsuperscript{***}              \\
    \texttt{rh}      & $\phantom{-}0.07$\phantom{\textsuperscript{***}}    & $0.12$\textsuperscript{**}\phantom{*}    & $\phantom{-}0.17$\textsuperscript{***}              & $0.21$\textsuperscript{**}\phantom{*}    \\
    \texttt{co}      & $\phantom{-}0.07$\phantom{\textsuperscript{***}}    & $0.02$\phantom{\textsuperscript{***}}    & $\phantom{-}0.36$\textsuperscript{***}              & $0.50$\textsuperscript{***}              \\
    \texttt{so2}     & $-0.06$\phantom{\textsuperscript{***}}              & $0.03$\phantom{\textsuperscript{***}}    & $\phantom{-}0.23$\textsuperscript{***}              & $0.24$\textsuperscript{***}              \\
    \texttt{no2}     & $\phantom{-}0.06$\phantom{\textsuperscript{***}}    & $0.04$\textsuperscript{*}\phantom{**}    & $-0.18$\textsuperscript{*}\phantom{**}              & $0.01$\phantom{\textsuperscript{***}} \\
    \texttt{hycarb}  & $\phantom{-}0.11$\phantom{\textsuperscript{***}}    & $0.05$\textsuperscript{*}\phantom{**}    & $\phantom{-}0.10$\phantom{\textsuperscript{***}}    & $0.02$\phantom{\textsuperscript{***}} \\
    \texttt{o3}      & $\phantom{-}0.03$\phantom{\textsuperscript{***}}    & $0.02$\phantom{\textsuperscript{***}}    & $-0.07$\phantom{\textsuperscript{***}}              & $0.09$\phantom{\textsuperscript{***}} \\
    \texttt{part}    & $\phantom{-}0.07$\phantom{\textsuperscript{***}}    & $0.07$\phantom{\textsuperscript{***}}    & $\phantom{-}0.02$\phantom{\textsuperscript{***}}    & $0.08$\phantom{\textsuperscript{***}} \\
    \texttt{tempr2}  & $-1.99$\textsuperscript{***}                        & $0.02$\phantom{\textsuperscript{***}}    & $-3.33$\textsuperscript{***}                        & $0.21$\phantom{\textsuperscript{***}} \\
    \hline
  \end{tabular}
\end{table}

The two methods agree on the leading drivers of the cardiovascular
factor: temperature, CO, SO$_2$, and humidity (\texttt{rh}) appear
as the four largest contributors in both columns
(SEM standardized coefficients $+3.81$, $+0.36$, $+0.23$, $+0.17$,
all at $p{<}0.001$;
NMF-FFB $\Theta_2$ $0.38$, $0.50$, $0.24$, $0.21$). The clear
methodological difference is in the temperature block, which is a
textbook multicollinearity symptom: because \texttt{tempr2} is the
squared standardized temperature, SEM assigns the two terms huge
opposing standardized coefficients, all at $p{<}0.001$
(cardio factor: $+3.81$ vs.\ $-3.33$;
\texttt{rmort}: $+2.32$ vs.\ $-1.99$), so each individual coefficient
is inflated and sign-unstable. NMF-FFB, constrained to be non-negative
and L1-regularized on $\Theta_2$, cannot absorb the collinearity
through a sign cancellation and instead distributes moderate
positive loads to \texttt{tempr} ($0.31$, $0.38$) and smaller loads
to \texttt{tempr2} ($0.02$, $0.21$). The same pattern shows up in
two smaller signals: SEM returns a counter-intuitive significant
negative coefficient for NO$_2$ on the cardiovascular factor
($-0.18$, $p{<}0.05$, likely an artifact of the
NO$_2$/CO/hydrocarbon combustion-channel collinearity), which
NMF-FFB shrinks to $0.01$ (bootstrap support rate $0.14$, n.s.,
independently corroborating the multicollinearity-artifact reading)
while keeping CO as the dominant proxy ($0.50$, $p{<}0.001$);
and SEM marks \texttt{rh}, \texttt{part}, and \texttt{hycarb}
non-significant on \texttt{rmort} while NMF-FFB retains them with
non-trivial weight ($0.12$, $0.07$, $0.05$). LA therefore exhibits
the multicollinearity-robustness of the non-negative, L1-regularized
formulation rather than a measurement-discovery advantage like
HS39: the strong correlation among regressors (\texttt{tempr} with
\texttt{tempr2}; CO with NO$_2$ and hydrocarbons) is what makes the
two methods diverge, and it is also what makes NMF-FFB's
representative-variable selection behavior visible.

\subsection{Mississippi health-risk analysis}
\label{subsec:mshealth}

\begin{figure}[htbp]
  \centering
  \includegraphics[width=0.98\linewidth]{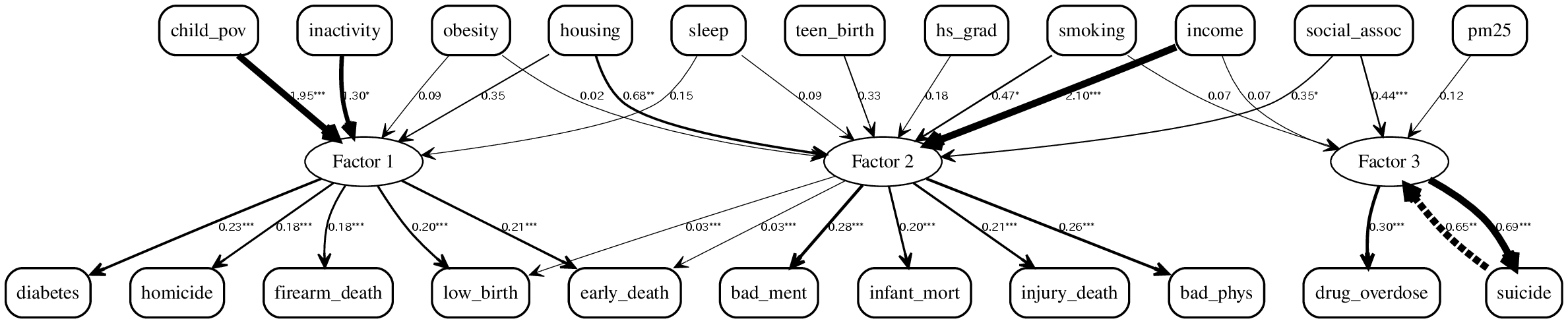}
  \caption{
    Path diagram estimated by NMF-FFB ($Q=3$) for Mississippi
    counties. Endogenous health outcomes $Y_1$:
    \texttt{early\_death} (premature death),
    \texttt{diabetes} (diabetes prevalence),
    \texttt{bad\_phys} (frequent physical distress),
    \texttt{suicide} (suicides),
    \texttt{bad\_ment} (frequent mental distress),
    \texttt{low\_birth} (low birth weight),
    \texttt{infant\_mort} (infant mortality),
    \texttt{injury\_death} (injury deaths),
    \texttt{homicide} (homicides),
    \texttt{drug\_overdose} (drug overdose deaths),
    \texttt{firearm\_death} (firearm fatalities);
    three latent factors; exogenous drivers $Y_2$:
    \texttt{child\_pov} (children in poverty),
    \texttt{income} (median household income; sign-reversed so that
    higher values indicate lower income),
    \texttt{smoking} (adult smoking),
    \texttt{inactivity} (physical inactivity),
    \texttt{obesity} (adult obesity),
    \texttt{hs\_grad} (high-school completion; sign-reversed so that
    higher values indicate lower completion),
    \texttt{sleep} (insufficient sleep),
    \texttt{pm25} (air pollution: particulate matter),
    \texttt{housing} (severe housing problems),
    \texttt{social\_assoc} (social associations; sign-reversed so that
    higher values indicate weaker association),
    \texttt{teen\_birth} (teen births).
    Edge convention: factor $\to Y_1$ edges encode $X$,
    $Y_1 \to$ factor edges encode $\Theta_1$ (latent feedback),
    and $Y_2 \to$ factor edges encode $\Theta_2$
    (exogenous channel).
    Factor~1 aggregates early-mortality and \texttt{diabetes}
    outcomes, Factor~2 aggregates distress and injury outcomes, and
    Factor~3 isolates \texttt{suicide} and \texttt{drug\_overdose}
    outcomes (``deaths of despair'').
    Under the CV-selected penalties, the only non-trivial
    latent-feedback pathway is the loading of the \texttt{suicide}
    outcome onto Factor~3 (value $0.65$, shown as the single return
    arrow in the diagram); all other $\Theta_1$ entries are zero to
    two decimals.
    Significance markers ($*$, $**$, $***$) on edge labels are
    one-sided support-rate tests at thresholds $0.95$, $0.99$,
    $0.999$ on $B_{\mathrm{boot}}{=}1000$ $X$-fixed bootstrap replicates
    (see \S\ref{subsec:feedback_metrics}).}
  \label{fig:mshealth_path}
\end{figure}

County-level data from the 2025 County Health Rankings (CHR)
\citep{CHR2025} were used.  
Eleven social determinants served as exogenous variables (poverty,
income, smoking, inactivity, obesity, education, sleep deprivation,
PM$_{2.5}$, housing problems, social association, and teen births), and
eleven health outcomes---including early mortality, diabetes, physical and
mental distress, infant outcomes, injury and homicide deaths, overdose,
suicide, and firearm deaths---were treated as endogenous variables.
All variables were rescaled to $[0,1]$, with protective indicators
\emph{sign-reversed} (multiplied by $-1$) to ensure that larger values
represent higher risk; this convention is used throughout
\S\ref{subsec:mshealth} and the corresponding figure caption
(Fig.~\ref{fig:mshealth_path}).

Cross-validation selected a latent dimension of $Q=3$. The estimated
path diagram is shown in Figure~\ref{fig:mshealth_path}.
The estimated feedback operator indicated modest latent feedback:
$\rho(X\Theta_1)\approx 0.446$ and $\mathrm{AR}\approx 1.025$, implying
a small amplification over direct effects.
The $95\%$ X-fixed bootstrap percentile intervals are
$[0.09,\,0.56]$ for $\rho(X\Theta_1)$ and $[1.00,\,1.09]$ for
$\mathrm{AR}$. The $\rho$ interval excludes zero, supporting
\emph{a modest but reliably positive latent feedback} at $N=35$
conditional on the data-driven measurement model;
the AR interval touches unity at its lower endpoint and therefore
supports only a small amplification over direct effects.

The estimated basis matrix $X$ (Table~\ref{tab:ms_basis}) yields three
interpretable latent components.  
Factor~1 loads on early mortality, diabetes, low birth weight,
homicide, and firearm death;
Factor~2 loads on mental and physical distress, injury deaths, and
infant mortality;
Factor~3 is concentrated on suicide, with a secondary loading on drug
overdose.

\begin{table}[htbp]
  \centering
  \caption{Estimated basis matrix $X$ for the Mississippi data
  ($Q=3$). Values shown as $0.00$ are penalized estimates at or near
  the non-negative boundary (see \S\ref{sec5} preamble).}
  \label{tab:ms_basis}
  \begin{tabular}{lccc}
    \hline
    Outcome & Factor~1 & Factor~2 & Factor~3 \\
    \hline
    \texttt{early\_death}      & 0.21 & 0.03 & 0.00 \\
    \texttt{diabetes}          & 0.23 & 0.00 & 0.00 \\
    \texttt{bad\_phys}         & 0.00 & 0.26 & 0.00 \\
    \texttt{suicide}           & 0.00 & 0.00 & 0.69 \\
    \texttt{bad\_ment}         & 0.00 & 0.28 & 0.01 \\
    \texttt{low\_birth}        & 0.20 & 0.03 & 0.00 \\
    \texttt{infant\_mort}      & 0.00 & 0.20 & 0.00 \\
    \texttt{injury\_death}     & 0.00 & 0.21 & 0.00 \\
    \texttt{homicide}          & 0.18 & 0.00 & 0.00 \\
    \texttt{drug\_overdose}    & 0.00 & 0.00 & 0.30 \\
    \texttt{firearm\_death}    & 0.18 & 0.00 & 0.00 \\
    \hline
  \end{tabular}
\end{table}

The latent feedback coefficients $\Theta_1^\top$ are extremely sparse
under the CV-selected penalties: only the suicide outcome carries a
non-trivial feedback loading onto Factor~3 (with value $0.65$),
while every other entry of $\Theta_1$ is zero to two decimal places.
In other words, the moderate $\hat\rho(X\Theta_1)\approx 0.45$ on
Mississippi is driven by a single self-reinforcing pathway through
suicide; drug overdose is connected to Factor~3 only through the
forward basis loading from drug overdose to Factor~3 in $X$ (value
$0.30$), not through a return arrow to the latent factor.
The exogenous pathways $\Theta_2$, reported alongside SEM in
Table~\ref{tab:ms_struct_compare}, indicate that Factor~1 is
associated with child poverty, inactivity, and housing problems;
Factor~2 mainly with income, smoking, housing, social association, and
teen birth, with low high-school graduation appearing as a smaller
non-significant positive entry; and Factor~3 mainly with social
association, with PM$_{2.5}$ appearing as a smaller non-significant
positive entry.


These latent pathways align with recent research emphasizing that 
``deaths of despair''---including suicide and drug overdose---are driven
primarily by social isolation and weakened community ties rather than by
material deprivation alone \citep{case2020}.  
NMF-FFB isolates this mechanism as a distinct latent component and
confirms the social-isolation channel at the bootstrap one-sided
$p{<}0.001$ level (\texttt{social\_assoc} $\to$ Factor 3:
$\Theta_2 = 0.44$; Table~\ref{tab:ms_struct_compare}). The two
principal economic-deprivation channels are likewise statistically
supported (\texttt{child\_pov} $\to$ Factor 1: $\Theta_2 = 1.95$ at
$p{<}0.01$ for the early-mortality cluster; \texttt{income} $\to$
Factor 2: $\Theta_2 = 2.10$ at $p{<}0.001$ for the distress/injury
cluster), and the single non-trivial latent-feedback path
(suicide outcome $\to$ Factor 3, $\Theta_1 = 0.65$) reaches bootstrap
support rate $0.991$ ($p{<}0.01$ in Fig.~\ref{fig:mshealth_path}).

The model exhibits high structural fidelity
(SC$_{\mathrm{map}}=0.950$, SC$_{\mathrm{cov}}=0.989$) and moderate
prediction error (MAE$=0.117$).
These results indicate that NMF-FFB provides a coherent non-negative
decomposition of county-level health outcomes and identifies distinct
exogenous pathways for each latent component, while latent feedback
remains limited given the small sample size ($N=35$).
The five-method comparison in Table~\ref{tab:real_comp} confirms that
NMF-FFB attains predictive accuracy in the same range as NMF-FF,
OLS-MIMIC, and the Standard SEM benchmark (reported via the BFGS
fallback after default lavaan ML fails to converge), while
uniquely retaining the equilibrium structure and the parts-based
basis interpretation. In the same Table~\ref{tab:real_comp},
removing all three penalties ($\lambda_X{=}\lambda_1{=}\lambda_2{=}0$)
drives $\hat\rho(X\Theta_1)\!\approx\!1.00$ (with AR$\approx 170$ and
MAE$=1.72$), so $X\Theta_1$ acquires an eigenvalue $\approx 1$ on the
column space of $Y_1$ and the model degenerates into the trivial
fixed point $Y_1\!\approx\!X\Theta_1 Y_1$.

\paragraph*{Reading the latent Leontief inverse.}
For Mississippi, the estimated Leontief inverse remains stable but is
close to feedforward: $\hat\rho(X\hat\Theta_1)\approx 0.44$ while
$\widehat{\mathrm{AR}}\approx 1.03$. Thus the higher-order terms in
$(I-X\hat\Theta_1)^{-1}$ are present but contribute only modestly beyond
the direct mapping $X\hat\Theta_2$. The practical reading is that the
county-level risk factors primarily act through direct latent pathways,
with only limited cumulative amplification among the latent health-risk
components.

\paragraph*{Comparison with standard SEM.}
For an external benchmark, we attempted a three-factor MIMIC fit on
the Mississippi dataset using the same NMF-FFB-informed grouping
and the same eleven exogenous variables. With $N=35$ against
approximately $50$ free ML parameters, the specification sits at
the boundary of identifiability: \texttt{lavaan} \citep{lavaan}
fails to converge under its default NLMINB optimizer and converges
to a statistically degenerate fixed point under BFGS
($\mathrm{CFI}{=}0.65$, $\mathrm{RMSEA}{=}0.22$, with standardized
coefficients exceeding $1$ in absolute value and standard errors up
to ${\sim}2{,}000$); \texttt{OpenMx} \citep{openmx} also converges
but with similarly poor fit ($\mathrm{CFI}{\approx}0.77$); the
\texttt{sem} package \citep{fox2006} fails outright on a
non-invertible model-implied covariance matrix. Only the
model-implied instrumental-variables 2SLS estimator
\citep[\texttt{MIIVsem};][]{bollen1996,bollen2021} fits the model
with usable standard errors. Table~\ref{tab:ms_struct_compare}
compares its structural coefficients with the NMF-FFB $\Theta_2$
entries.
This failure mode is expected for a covariance-structure ML fit in
which roughly $50$ free parameters must be supported by only $35$
counties: the likelihood surface is nearly flat or ill-conditioned in
several directions, and numerical optimization depends on Hessian-like
curvature information that becomes unreliable. NMF-FFB does not remove
the inferential limitations of $N=35$, but it avoids this particular
failure mode by fitting a penalized Frobenius reconstruction objective
through multiplicative block updates under non-negativity and
regularization, without inverting a sample covariance matrix or requiring
a positive-definite ML information matrix.

\begin{table}[htbp]
  \centering\scriptsize
  \caption{Mississippi structural coefficients: instrumental-variable
  SEM (MIIV-2SLS, signed) vs.\ NMF-FFB. SEM coefficients are
  unstandardized on the same $[0,1]$-rescaled data;
  significance markers as in
  Table~\ref{tab:hs39_struct_compare}. The
  suicide-factor block of the SEM fit is affected by a Heywood case
  (one negative latent variance estimate), so its coefficients should
  be read with caution.}
  \label{tab:ms_struct_compare}
  \begin{tabular}{l@{\hskip 6pt}r@{\hskip 6pt}r@{\hskip 14pt}r@{\hskip 6pt}r@{\hskip 14pt}r@{\hskip 6pt}r}
    \hline
                  & \multicolumn{2}{c}{Factor 1} & \multicolumn{2}{c}{Factor 2} & \multicolumn{2}{c}{Factor 3} \\
                  & \multicolumn{2}{c}{(early-mortality)} & \multicolumn{2}{c}{(distress/injury)} & \multicolumn{2}{c}{(despair)} \\
    Variable      & \multicolumn{1}{c}{SEM} & \multicolumn{1}{c}{NMF-FFB} & \multicolumn{1}{c}{SEM} & \multicolumn{1}{c}{NMF-FFB} & \multicolumn{1}{c}{SEM} & \multicolumn{1}{c}{NMF-FFB} \\
    \hline
    \texttt{child\_pov}    & $\phantom{-}0.51$\textsuperscript{*}\phantom{**}    & $1.95$\textsuperscript{**}\phantom{*}    & $-0.15$\phantom{\textsuperscript{***}}              & $0.00$\phantom{\textsuperscript{***}} & $-0.26$\phantom{\textsuperscript{***}}              & $0.00$\phantom{\textsuperscript{***}} \\
    \texttt{income}        & $-0.02$\phantom{\textsuperscript{***}}              & $0.00$\phantom{\textsuperscript{***}}    & $\phantom{-}0.57$\textsuperscript{***}              & $2.10$\textsuperscript{***}              & $\phantom{-}0.72$\textsuperscript{*}\phantom{**}    & $0.07$\phantom{\textsuperscript{***}} \\
    \texttt{smoking}       & $\phantom{-}0.34$\phantom{\textsuperscript{***}}    & $0.00$\phantom{\textsuperscript{***}}    & $\phantom{-}0.32$\phantom{\textsuperscript{***}}    & $0.47$\textsuperscript{*}\phantom{**}    & $\phantom{-}0.67$\phantom{\textsuperscript{***}}    & $0.07$\phantom{\textsuperscript{***}} \\
    \texttt{inactivity}    & $-0.08$\phantom{\textsuperscript{***}}              & $1.30$\textsuperscript{*}\phantom{**}    & $-0.22$\phantom{\textsuperscript{***}}              & $0.00$\phantom{\textsuperscript{***}} & $-0.44$\phantom{\textsuperscript{***}}              & $0.00$\phantom{\textsuperscript{***}} \\
    \texttt{obesity}       & $\phantom{-}0.16$\phantom{\textsuperscript{***}}    & $0.09$\phantom{\textsuperscript{***}}    & $-0.02$\phantom{\textsuperscript{***}}              & $0.02$\phantom{\textsuperscript{***}} & $-0.34$\phantom{\textsuperscript{***}}              & $0.00$\phantom{\textsuperscript{***}} \\
    \texttt{hs\_grad}      & $-0.09$\phantom{\textsuperscript{***}}              & $0.00$\phantom{\textsuperscript{***}}    & $\phantom{-}0.16$\phantom{\textsuperscript{***}}    & $0.18$\phantom{\textsuperscript{***}} & $-0.62$\textsuperscript{**}\phantom{*}              & $0.00$\phantom{\textsuperscript{***}} \\
    \texttt{sleep}         & $-0.08$\phantom{\textsuperscript{***}}              & $0.15$\phantom{\textsuperscript{***}}    & $-0.04$\phantom{\textsuperscript{***}}              & $0.09$\phantom{\textsuperscript{***}} & $-0.57$\textsuperscript{**}\phantom{*}              & $0.00$\phantom{\textsuperscript{***}} \\
    \texttt{pm25}          & $\phantom{-}0.06$\phantom{\textsuperscript{***}}    & $0.00$\phantom{\textsuperscript{***}}    & $-0.09$\phantom{\textsuperscript{***}}              & $0.00$\phantom{\textsuperscript{***}} & $-0.10$\phantom{\textsuperscript{***}}              & $0.12$\phantom{\textsuperscript{***}} \\
    \texttt{housing}       & $\phantom{-}0.23$\textsuperscript{.}\phantom{**}    & $0.35$\phantom{\textsuperscript{***}}    & $\phantom{-}0.16$\phantom{\textsuperscript{***}}    & $0.68$\textsuperscript{**}\phantom{*}    & $\phantom{-}0.06$\phantom{\textsuperscript{***}}    & $0.00$\phantom{\textsuperscript{***}} \\
    \texttt{social\_assoc} & $\phantom{-}0.12$\phantom{\textsuperscript{***}}    & $0.00$\phantom{\textsuperscript{***}}    & $\phantom{-}0.07$\phantom{\textsuperscript{***}}    & $0.35$\textsuperscript{*}\phantom{**}    & $\phantom{-}0.27$\textsuperscript{.}\phantom{**}    & $0.44$\textsuperscript{***}              \\
    \texttt{teen\_birth}   & $\phantom{-}0.20$\phantom{\textsuperscript{***}}    & $0.00$\phantom{\textsuperscript{***}}    & $\phantom{-}0.32$\textsuperscript{*}\phantom{**}    & $0.33$\phantom{\textsuperscript{***}} & $\phantom{-}0.51$\textsuperscript{.}\phantom{**}    & $0.00$\phantom{\textsuperscript{***}} \\
    \hline
  \end{tabular}
\end{table}

The two methods agree on the lead pathways: child poverty drives
the mortality factor (SEM $+0.51$, $p{<}0.05$, vs.\ NMF-FFB
$\Theta_2{=}1.95$); low income drives the distress factor
($+0.57$, $p{<}0.001$, vs.\ $2.10$); \texttt{teen\_birth}
contributes secondarily to distress ($+0.32$, $p{<}0.05$,
vs.\ $0.33$, identical to two decimals); and \texttt{social\_assoc}
contributes to the suicide factor ($+0.27$, $p{<}0.1$, vs.\ $0.44$).
Because the MIIV-2SLS suicide-factor block is affected by a Heywood
case (Table~\ref{tab:ms_struct_compare}), the following signed SEM
coefficients are used diagnostically rather than as a clean substantive
SEM solution.
The counter-intuitive significant negatives on the suicide
factor---\texttt{hs\_grad} $-0.62$ ($p{<}0.01$) and
\texttt{sleep} $-0.57$ ($p{<}0.01$)---are best read as
multicollinearity artifacts. After the sign flip described in
\S\ref{subsec:mshealth}, \texttt{hs\_grad} encodes low high-school
completion and \texttt{sleep} encodes insufficient-sleep prevalence,
so both variables are risk indicators for which positive signs
would be epidemiologically expected.
The eleven exogenous variables, however, exhibit $27$ of $55$ pairs
with $|r|\!\ge\!0.5$ (max $r{=}0.90$ for
\texttt{smoking}--\texttt{inactivity}) and severe variance-inflation
factors (\texttt{inactivity} $18$, \texttt{smoking} $13$,
\texttt{child\_pov} $11$). Just as for tempr/tempr$^2$ in LA
(\S\ref{subsec:pollution}), the unrestricted SEM concentrates the
suicide signal on \texttt{income} and assigns offsetting negative
coefficients to the variables that share information with it, while
NMF-FFB's non-negative L1-regularized $\Theta_2$ shrinks the
suicide-factor coefficients on both \texttt{hs\_grad} and
\texttt{sleep} to zero. The bootstrap on NMF-FFB independently
corroborates this multicollinearity-artifact reading: both
\texttt{hs\_grad} and \texttt{sleep} on Factor 3 are
bootstrap-non-significant (support rates $0.001$ and $0.04$,
respectively), confirming that the L1 shrinkage to zero is not
an L1 idiosyncrasy but reflects genuine non-identifiability of
these signed signals under the observed collinearity structure.
For Mississippi, the methodological payoff of NMF-FFB is therefore
that it delivers a stable, interpretable solution under CV-selected
penalties where ML-based SEM either fails or converges to
degenerate fixed points. The point is not that NMF-FFB is uniformly more
accurate than SEM, but that it returns an auditable exploratory
decomposition for a small, public county-health dataset in which a clean
covariance-based SEM benchmark is numerically fragile and requires
dataset-specific restrictions. This is precisely the kind of setting in
which CHR-style public data are scientifically useful but difficult to
analyze with a fully confirmatory covariance-SEM workflow. NMF-FFB also
produces $\Theta_1$ and $(I-X\Theta_1)^{-1}$, so the analysis retains a
latent-feedback and cumulative-effect layer that the usable standard-SEM
fallback does not provide.

\section{Discussion and conclusion}\label{sec6}

Empirically, the three applications spanned the regimes the framework
is designed to distinguish: HS39 displayed moderate latent
amplification, the LA pollution--mortality system was effectively
feedforward, and the Mississippi health-risk data exhibited moderate
$\rho(X\Theta_1)$ but only small cumulative amplification. Because the
Mississippi dataset is small ($N=35$),
its structural conclusions rest on equilibrium metrics and
bootstrap percentiles rather than individual parameter estimates;
the bootstrap intervals for $\rho(X\Theta_1)$ and AR were narrow on
HS39 and LA but appreciably wider on Mississippi, and the
corresponding $\hat\rho$ and AR should be read as
regularization-dependent exploratory diagnostics rather than as
scale-free population estimators (cf.\ \S\ref{subsubsec:Nscale}).
The $X$-fixed bootstrap of \S\ref{subsec:feedback_metrics}
additionally computes entry-level percentile CIs as an internal
consistency check and displays one-sided support-rate significance
markers for the non-negative path coefficients $\Theta_1$ and
$\Theta_2$, allowing NMF-FFB to be displayed alongside standard SEM
path coefficients while recognizing that the inferential basis is a
one-sided fixed-$X$ bootstrap rather than a likelihood-based Wald test.
This path-level inference layer was absent from the originally
submitted manuscript.
SC$_{\mathrm{map}}$ was high for HS39, moderate for LA, and at the
pre-specified acceptability boundary for Mississippi
($0.999, 0.882, 0.950$), while SC$_{\mathrm{cov}}$ exceeded $0.98$ in
all three applications. Thus the equilibrium operator
$M_{\mathrm{model}}$ retains substantial, dataset-dependent alignment
with the feedforward baseline and reproduces the observed
second-moment structure even though $(\Theta_1,\Theta_2)$ themselves
are not point-identified.

\medskip\noindent
Several limitations warrant mention.
First, $(\Theta_1,\Theta_2)$ are not uniquely identifiable; emphasis
should therefore be placed on equilibrium behavior rather than on the
individual parameter values.
Second, interpretable solutions often require strong regularization.
Third (a), $\hat\rho$ and AR are regularization-dependent diagnostics
rather than scale-free population estimators: their numerical values
shift with $(\lambda_X,\lambda_1,\lambda_2)$ and, under any fixed
penalty, with $N$.
Third (b), as a consequence, absolute magnitudes of $\hat\rho$ or AR
should not be compared across datasets of different $N$ unless the
$N$-scaled convention of Section~\ref{subsubsec:Nscale} is applied;
within a single dataset the fixed-$\lambda$ reading remains
interpretable as an ordinal descriptor of feedback strength.
Fourth, the current formulation assumes a static equilibrium and does
not model temporal evolution.
Fifth, although NMF-FFB only requires that $(Y_1,Y_2)$ be non-negative
and does not formally require min-max rescaling to $[0,1]$, we
recommend min-max rescaling $(x-\min)/(\max-\min)$ as the default
preprocessing step for the two estimator-fairness reasons stated in
\S\ref{subsec:SCcov}: it makes every variable contribute on the same
scale to the Frobenius reconstruction loss, and it makes the
$\ell_1$ penalties on $\Theta_1$ and $\Theta_2$ act with uniform
strength across variables of $Y_1$ and $Y_2$. The main caveat is
its sensitivity to outliers: a single extreme value compresses
the remaining observations toward zero. Practitioners analyzing
heavy-tailed or outlier-prone data should consider \emph{robust}
alternatives---winsorization at, e.g., the $1\%$/$99\%$ quantiles
before rescaling, or quantile-based (rank-preserving) transforms
into $[0,1]$---as outlier-resistant replacements. For the three
datasets analyzed in this paper, a sensitivity check in which
entries below the $1$st or above the $99$th percentile were trimmed
before min-max rescaling did not change the qualitative conclusions
about feedback strength or input--output fidelity.
Sixth, the recommended $\lambda_X{=}100$ is calibrated for
$[0,1]$-rescaled data; under other scales or substantially different
$N$ it should be re-tuned per the off-ratio criterion of
\S\ref{subsec:ortho_check} (see \S\ref{subsec:crossvalidation} for
the full rule).

Future work includes (i) dynamic extensions incorporating time-series
structure, uncertainty quantification for equilibrium effects, and
hybrid models integrating graphical or network constraints; (ii)
applications to large-scale systems in which intermediate flows are
unobserved; and (iii) fully data-driven tuning of $(\lambda_1,
\lambda_2)$ via empirical-Bayes or automatic-relevance-determination
\citep{TanFevotte2013, Cemgil2009, schmidt2009} as an alternative to
the cross-validation grids used here, which were chosen to let the
analyst encode the differing a-priori sparsity expectations of
$\Theta_1$ and $\Theta_2$ through separate penalties.

\medskip
\noindent\textit{Software availability.}
The stable public release of the \texttt{nmfkc} R package
\citep{satoh2025nmfkc} is available on CRAN (version~0.6.7,
2026-04-15; \url{https://CRAN.R-project.org/package=nmfkc};
\url{https://doi.org/10.32614/CRAN.package.nmfkc}). Some functions
used for the analyses reported in this paper were added after the
CRAN release and are currently available from the GitHub main branch,
with package documentation at \url{https://ksatohds.github.io/nmfkc/}.

\section*{Acknowledgements}

The author thanks the anonymous reviewers for their constructive
comments and insightful suggestions, which helped improve the clarity
and quality of this paper.

\section*{Statements and Declarations}

\noindent\textbf{Funding.} This work was partly supported by JSPS
KAKENHI Grant Numbers 22K11930, 25K15229, 24K03007, 25H00482, and a
research grant from the Fuji Seal Foundation.

\noindent\textbf{Competing Interests.} The author states that there is
no conflict of interest.

\noindent\textbf{Ethical Approval.} Not applicable.

\noindent\textbf{Data Availability.} The datasets generated or
analyzed during the current study are available from the corresponding
author on reasonable request.

\noindent\textbf{Code Availability.} The stable public release of
the R package \texttt{nmfkc} \citep{satoh2025nmfkc} is available on
CRAN at \url{https://CRAN.R-project.org/package=nmfkc}
(\url{https://doi.org/10.32614/CRAN.package.nmfkc}). Some functions
used in this paper are currently available from the GitHub main
branch and documented at \url{https://ksatohds.github.io/nmfkc/}.

\begin{appendices}
\section{Derivation of multiplicative updates}\label{app:derivation}

This appendix provides a self-contained derivation of the multiplicative
updates for NMF-FFB introduced in Section~\ref{subsec:updates}. The
reconstruction part follows the auxiliary-function construction of Lee
and Seung \citep{lee1999,lee2000}; the regularized terms are incorporated
through the corresponding positive-gradient/KKT splits. We also document
the convergence criterion used in the implementation. Throughout,
$Y_1\in\mathbb{R}_{+}^{P_1\times N}$
is the endogenous block, $Y_2\in\mathbb{R}_{+}^{P_2\times N}$ is the
exogenous block, $X\in\mathbb{R}_{+}^{P_1\times Q}$ is the basis,
$\Theta_1\in\mathbb{R}_{+}^{Q\times P_1}$ and
$\Theta_2\in\mathbb{R}_{+}^{Q\times P_2}$ are the structural-coefficient
blocks, and we write
\[
B \;=\; \Theta_1 Y_1 + \Theta_2 Y_2,\qquad
\widehat{Y}_1 \;=\; XB,
\]
so that NMF-FFB fits $Y_1 \approx \widehat{Y}_1$.  The penalized
squared-error objective of Section~\ref{subsec:regularization} is
\begin{align}
L(X,\Theta_1,\Theta_2)
&= \tfrac12\,\|Y_1 - XB\|_F^{2}
 \;+\; \tfrac{\lambda_X}{2}\,\|X^\top X - \mathrm{diag}(X^\top X)\|_F^{2}\notag\\
&\qquad\;+\; \lambda_1\|\Theta_1\|_1 \;+\; \lambda_2\|\Theta_2\|_1.
\label{eq:app:L}
\end{align}
The $1/2$ factor on the reconstruction term is adopted in the derivation
to keep gradients clean; the implementation reports
$\|Y_1-XB\|_F^{2}$ (without the $1/2$), which only rescales the
effective $\ell_1$ penalty by a factor of two and does not affect the
form of the updates.

\subsection{The auxiliary-function method}\label{app:aux}

Let $\theta$ denote any one of the block variables $X$, $\Theta_1$, or
$\Theta_2$, with the remaining blocks held fixed.  A function
$G(\theta,\tilde\theta)$ is an \emph{auxiliary function} for $L(\theta)$
at $\tilde\theta$ if
\begin{equation}
G(\theta,\theta) \;=\; L(\theta)\qquad\text{and}\qquad
G(\theta,\tilde\theta) \;\ge\; L(\theta)\quad\text{for all } \theta\ge 0.
\label{eq:app:aux-def}
\end{equation}
Given an auxiliary function, the update
\(
\theta^{(t+1)} \in \arg\min_{\theta\ge 0} G(\theta,\theta^{(t)})
\)
satisfies the descent property
\[
L\!\left(\theta^{(t+1)}\right)
\;\le\; G\!\left(\theta^{(t+1)},\theta^{(t)}\right)
\;\le\; G\!\left(\theta^{(t)},\theta^{(t)}\right)
\;=\; L\!\left(\theta^{(t)}\right).
\]
Lee and Seung \citep{lee2000} showed that, for the unpenalized
squared-error NMF loss, Jensen's inequality applied to the quadratic
term $\tfrac12\|Y - XB\|_F^{2}$ produces a \emph{separable} quadratic
upper bound whose block-wise minimizer reduces to a closed-form ratio of
non-negative quantities --- namely, the classical multiplicative update.
The same device applies to the reconstruction part of each block of
NMF-FFB; the substitution
$B=\Theta_1 Y_1 + \Theta_2 Y_2$ is linear in each of
$\Theta_1,\Theta_2$. The regularization terms are then added through the
positive part of the KKT gradient split in the denominators. See
Cichocki~et~al.\ \citep{cichocki2009} for a
comprehensive treatment of the auxiliary-function machinery in general
NMF.

\subsection{Update for the basis \texorpdfstring{$X$}{X}}\label{app:X}

Fix $(\Theta_1,\Theta_2)$, equivalently $B$.  The $X$-gradient of the
reconstruction term is
\[
\nabla_{X}\,\tfrac12\|Y_1-XB\|_F^{2}
\;=\; -Y_1 B^\top + XBB^\top
\;=\; -\underbrace{Y_1 B^\top}_{=:\,\nabla^{-}_X}
      +\underbrace{XBB^\top}_{=:\,\nabla^{+}_X},
\]
where both $\nabla^{-}_X$ and $\nabla^{+}_X$ are element-wise
non-negative.  The orthogonality penalty
$P_X(X)=\tfrac{\lambda_X}{2}\|X^\top X - \mathrm{diag}(X^\top X)\|_F^{2}$
has gradient
\begin{equation}
\nabla_{X} P_X(X)
\;=\; 2\lambda_X\,X\bigl(X^\top X - \mathrm{diag}(X^\top X)\bigr),
\label{eq:app:ortho-grad}
\end{equation}
which is element-wise non-negative because all entries of $X$ are
non-negative and the off-diagonal entries of $X^\top X$ are non-negative
inner products.  Since $\lambda_X$ is a user-chosen hyperparameter, the
factor of $2$ in \eqref{eq:app:ortho-grad} is absorbed into $\lambda_X$
by the reparametrization $2\lambda_X \mapsto \lambda_X$, and we use this
convention throughout: when the gradient is plugged into the
multiplicative update \eqref{eq:app:Xupdate} below, the coefficient is
written simply as $\lambda_X$.  This is standard in the NMF literature
(e.g.\ Ding et al.\ \citep{ding2006}).

For the $X$-update, Lee and Seung's auxiliary-function construction
\citep[Eq.~(14)]{lee2000} provides the separable quadratic upper bound
\[
\begin{aligned}
G_X(X,\tilde X)
\;=\;{}& F(\tilde X)
      \;+\; \operatorname{tr}\!\bigl[(X-\tilde X)^{\top}\nabla F(\tilde X)\bigr] \\
      &\;+\; \tfrac12 \sum_{p,q}
         \frac{(\tilde X B B^\top)_{pq}}{\tilde X_{pq}}\,
         \bigl(X_{pq}-\tilde X_{pq}\bigr)^{2},
\end{aligned}
\]
where $F(X)=\tfrac12\|Y_1 - XB\|_F^{2}$ and
$\nabla F(\tilde X) = -Y_1 B^\top + \tilde X BB^\top$.  The diagonal
``curvature'' coefficient $(\tilde X BB^\top)_{pq}/\tilde X_{pq}$
dominates the Hessian $BB^\top$ in the sense of Lee--Seung
\citep[Lemma~2]{lee2000}, so
$G_X(X,\tilde X)\ge F(X)$ and $G_X(\tilde X,\tilde X)=F(\tilde X)$,
satisfying \eqref{eq:app:aux-def}. The orthogonality penalty is then
incorporated by adding the non-negative positive-gradient component
$\lambda_X\,X(X^\top X-\mathrm{diag}(X^\top X))$ to the denominator,
following the standard KKT-splitting form used in orthogonal NMF
\citep{ding2006,cichocki2009}. This step should be read as a
regularized multiplicative split, not as a separate global
auxiliary-function proof for the full orthogonality penalty.
Differentiating $G_X$ entrywise and solving $\partial G_X/\partial X_{pq}=0$
gives the closed-form stationary point
$X_{pq} = \tilde X_{pq}(Y_1 B^\top)_{pq}/(\tilde X BB^\top)_{pq}$,
which, after including the orthogonality-penalty positive-gradient term
in the denominator and applying the rescaling convention above, yields the
Karush--Kuhn--Tucker (KKT) ratio
\begin{equation}
\boxed{\;
X \;\longleftarrow\; X \,\odot\,
\frac{Y_1 B^\top}
     {XBB^\top \;+\; \lambda_X\, X\bigl(X^\top X - \mathrm{diag}(X^\top X)\bigr)}\;}
\label{eq:app:Xupdate}
\end{equation}
with elementwise multiplication and division.  After every $X$-update we
renormalize the columns of $X$ so that $\mathbf{1}^\top X = \mathbf{1}^\top$
(sum-to-one constraint), which removes the residual scale ambiguity
inherent in the factorization $Y_1\approx XB$.
This renormalization is part of the implementation, but it is not itself
the minimizer of the auxiliary function; the formal descent statement
below therefore applies to the raw multiplicative block updates prior to
this gauge-fixing rescaling.

\subsection{Update for \texorpdfstring{$\Theta_1$}{Theta1}}\label{app:Theta1}

Now fix $X$ and $\Theta_2$.  Because $B$ depends linearly on $\Theta_1$
via $B=\Theta_1 Y_1 + \Theta_2 Y_2$ and $\widehat{Y}_1 = XB$, the chain
rule gives
\begin{align}
\nabla_{\Theta_1}\,\tfrac12\|Y_1-XB\|_F^{2}
&= -X^\top (Y_1 - XB)\,Y_1^\top \notag\\
&= -X^\top Y_1 Y_1^\top + X^\top \widehat{Y}_1 Y_1^\top.
\label{eq:app:Theta1-grad}
\end{align}
Both terms are non-negative because $X, Y_1, \widehat{Y}_1$ are
non-negative.  The subgradient of $\lambda_1\|\Theta_1\|_1$ on the
positive orthant is the all-ones matrix $\lambda_1\mathbf{1}$ of size
$Q\times P_1$.

To derive the multiplicative update for $\Theta_1$ explicitly, note
that for fixed $X$ and $\Theta_2$ the residual is quadratic in
$\Theta_1$, with
$(XB)_{ij}
 = \sum_{q,k} X_{iq}(\Theta_1)_{qk}(Y_1)_{kj}
 + \sum_{q,k} X_{iq}(\Theta_2)_{qk}(Y_2)_{kj}$.
Applying Jensen's inequality with the Lee--Seung auxiliary-function
weights
$\alpha_{ijqk}
 = X_{iq}(\tilde\Theta_1)_{qk}(Y_1)_{kj}\big/(X\tilde B)_{ij}\ge 0$
(satisfying $\sum_{q,k}\alpha_{ijqk}+\sum_{q,k}\alpha^{(2)}_{ijqk}=1$,
where $\alpha^{(2)}$ is the analogous weight for the $\Theta_2 Y_2$
component; \citealp{lee2000}) and convexity of $x\mapsto x^{2}$ yields
the separable quadratic upper bound
\begin{align*}
G_{\Theta_1}(\Theta_1,\tilde\Theta_1)
&= F(\tilde\Theta_1)
 + \operatorname{tr}\!\bigl[(\Theta_1-\tilde\Theta_1)^{\top}
                            \nabla_{\Theta_1} F(\tilde\Theta_1)\bigr] \\
&\quad + \tfrac12\!\sum_{q,k}\!
   \frac{(X^\top \widehat{Y}_1 Y_1^\top)_{qk}}{(\tilde\Theta_1)_{qk}}
   \bigl((\Theta_1)_{qk}-(\tilde\Theta_1)_{qk}\bigr)^{2},
\end{align*}
which satisfies $G_{\Theta_1}(\Theta_1,\tilde\Theta_1)\ge F(\Theta_1)$
and $G_{\Theta_1}(\tilde\Theta_1,\tilde\Theta_1)=F(\tilde\Theta_1)$.
Adding the $\ell_1$ subgradient contribution $\lambda_1\mathbf{1}$ to
the linear coefficient and solving
$\partial G_{\Theta_1}/\partial(\Theta_1)_{qk}=0$ with
\eqref{eq:app:Theta1-grad} gives the KKT ratio
\begin{equation}
\boxed{\;
\Theta_1 \;\longleftarrow\; \Theta_1 \,\odot\,
\frac{X^\top Y_1 Y_1^\top}
     {X^\top \widehat{Y}_1 Y_1^\top \;+\; \lambda_1}\;}
\label{eq:app:Theta1update}
\end{equation}
where $\lambda_1$ is broadcast to a $Q\times P_1$ all-ones matrix
scaled by $\lambda_1$.

\subsection{Update for \texorpdfstring{$\Theta_2$}{Theta2}}\label{app:Theta2}

Replacing $Y_1^\top$ by $Y_2^\top$ in \eqref{eq:app:Theta1-grad} gives
\[
\nabla_{\Theta_2}\,\tfrac12\|Y_1-XB\|_F^{2}
\;=\; -X^\top Y_1 Y_2^\top + X^\top \widehat{Y}_1 Y_2^\top,
\]
and the same majorization argument yields
\begin{equation}
\boxed{\;
\Theta_2 \;\longleftarrow\; \Theta_2 \,\odot\,
\frac{X^\top Y_1 Y_2^\top}
     {X^\top \widehat{Y}_1 Y_2^\top \;+\; \lambda_2}\;}
\label{eq:app:Theta2update}
\end{equation}
with $\lambda_2$ broadcast to the $Q\times P_2$ all-ones matrix scaled
by $\lambda_2$.  Note that the $\Theta_2$-update differs structurally
from $\Theta_1$ only through the exogenous factor $Y_2$: because $Y_2$
does \emph{not} appear on the left-hand side of
$Y_1\approx X(\Theta_1 Y_1+\Theta_2 Y_2)$, there is no feedback through
$\widehat{Y}_1$ to $Y_2$ in the gradient.  This is the algebraic origin
of the asymmetry between the endogenous block $Y_1$ and the exogenous
block $Y_2$.

\subsection{Convergence criterion and descent caveat}\label{app:conv}

\paragraph*{Descent scope.}
As shown in \S\ref{app:aux}, each block update
\eqref{eq:app:Xupdate}, \eqref{eq:app:Theta1update},
\eqref{eq:app:Theta2update} contains the Lee--Seung auxiliary-function
step for the reconstruction loss. For the unpenalized raw block updates,
this gives descent of
$F_{\mathrm{rec}}(X,\Theta_1,\Theta_2)=\tfrac12\|Y_1-XB\|_F^2$:
\[
F_{\mathrm{rec}}\!\left(X^{(t+1)},\Theta_1^{(t+1)},\Theta_2^{(t+1)}\right)
\;\le\;
F_{\mathrm{rec}}\!\left(X^{(t)},\Theta_1^{(t)},\Theta_2^{(t)}\right).
\]
For the penalized algorithm used here, the $\ell_1$ and orthogonality
terms are incorporated by KKT-type positive-gradient splitting, and each
sweep additionally renormalizes the columns of $X$ to fix the scale gauge
of the factorization. We therefore do not claim a global monotone
decrease theorem for the fully regularized and renormalized sweep.
Instead, the auxiliary-function argument justifies the reconstruction
part of the multiplicative steps, the KKT split gives the stated
regularized ratios, and the stopping rule below monitors the practical
convergence of the implemented iteration.

\paragraph*{Caveat: non-convexity.}
$L$ is not jointly convex in $(X,\Theta_1,\Theta_2)$ --- neither in
standard NMF nor in its NMF-FFB extension. Under the usual
multiplicative-update regularity conditions, one should expect at most
convergence to a local stationary point, not to the global minimum; for
the fully regularized and renormalized implementation, this is treated as
a practical convergence criterion rather than a global theorem.
To mitigate sensitivity to the starting value we initialize via the
feedforward NNDSVDar scheme described in Section~\ref{subsec:init},
which is deterministic and has been reported to yield near-global
optima for standard NMF \citep{boutsidis2008}.

\paragraph*{Stopping rule.}
Let $L^{(t)}$ denote the value of \eqref{eq:app:L} after iteration $t$.
The implementation terminates when
\begin{equation}
\frac{\bigl|L^{(t)} - L^{(t-1)}\bigr|}
     {\max\!\bigl(\bigl|L^{(t)}\bigr|,\,1\bigr)}
\;\le\; \varepsilon,
\qquad
\varepsilon = 10^{-6},
\label{eq:app:stop}
\end{equation}
holds, or when a maximum iteration count $t_{\max}=5000$ is reached.

\end{appendices}

\bibliography{ksatoh2025}

\begin{thebibliography}{37}
\providecommand{\natexlab}[1]{#1}
\providecommand{\url}[1]{{#1}}
\providecommand{\urlprefix}{URL }
\providecommand{\doi}[1]{\url{https://doi.org/#1}}
\providecommand{\eprint}[2][]{\url{#2}}
 \bibcommenthead

\bibitem[{Bollen(1989)}]{bollen1989}
Bollen KA (1989) Structural equations with latent variables, vol~25. Wiley

\bibitem[{Bollen(1996)}]{bollen1996}
Bollen KA (1996) An alternative two stage least squares (2{SLS}) estimator for
  latent variable equations. Psychometrika 61(1):109--121.
  \doi{10.1007/BF02296961}

\bibitem[{Bollen et~al(2021)Bollen, Fisher, Giordano, Lilly, Luo, and
  Ye}]{bollen2021}
Bollen KA, Fisher Z, Giordano ML, et~al (2021) An introduction to model implied
  instrumental variables using two stage least squares ({MIIV-2SLS}) in
  structural equation models ({SEMs}). Psychological Methods 27(5):752--772.
  \doi{10.1037/met0000297}

\bibitem[{Boutsidis and Gallopoulos(2008)}]{boutsidis2008}
Boutsidis C, Gallopoulos E (2008) Svd based initialization: A head start for
  nonnegative matrix factorization. Pattern recognition 41(4):1350--1362

\bibitem[{Brook et~al(2010)Brook, Rajagopalan, Pope~III, Brook, Bhatnagar,
  Diez-Roux, Holguin, Hong, Luepker, Mittleman et~al}]{brook2010}
Brook RD, Rajagopalan S, Pope~III CA, et~al (2010) Particulate matter air
  pollution and cardiovascular disease: an update to the scientific statement
  from the american heart association. Circulation 121(21):2331--2378

\bibitem[{Cai et~al(2023)Cai, Gu, and Kenney}]{cai2023}
Cai Y, Gu H, Kenney T (2023) Rank selection for non-negative matrix
  factorization. Statistics in Medicine 42(30):5676--5693.
  \doi{10.1002/sim.9934}

\bibitem[{Case and Deaton(2020)}]{case2020}
Case A, Deaton A (2020) Deaths of Despair and the Future of Capitalism.
  Princeton University Press,
  \urlprefix\url{http://www.jstor.org/stable/j.ctvpr7rb2}

\bibitem[{Cemgil(2009)}]{Cemgil2009}
Cemgil AT (2009) Bayesian inference for nonnegative matrix factorisation
  models. Computational Intelligence and Neuroscience 2009:Article ID 785152.
  \doi{10.1155/2009/785152}, pMID: 19536273

\bibitem[{Cichocki et~al(2009)Cichocki, Zdunek, Phan, and Amari}]{cichocki2009}
Cichocki A, Zdunek R, Phan A, et~al (2009) Nonnegative Matrix and Tensor
  Factorizations: Applications to Exploratory Multi-way Data Analysis and Blind
  Source Separation. Wiley

\bibitem[{Ding et~al(2006)Ding, Li, Peng, and Park}]{ding2006}
Ding C, Li T, Peng W, et~al (2006) Orthogonal nonnegative matrix
  t-factorizations for clustering. In: Proceedings of the 12th ACM SIGKDD
  international conference on Knowledge discovery and data mining, pp 126--135

\bibitem[{Donoho and Stodden(2003)}]{Donoho2003}
Donoho D, Stodden V (2003) When does non-negative matrix factorization give a
  correct decomposition into parts? In: Thrun S, Saul L, Sch{\"o}lkopf B (eds)
  Advances in Neural Information Processing Systems, vol~16. MIT Press, pp
  1141--1148

\bibitem[{Fox(2006)}]{fox2006}
Fox J (2006) Structural equation modeling with the {sem} package in {R}.
  Structural Equation Modeling: A Multidisciplinary Journal 13(3):465--486.
  \doi{10.1207/s15328007sem1303\_7}

\bibitem[{Galeano et~al(2024)Galeano, Santos, Jimenez, and
  Paccanaro}]{galeano2024}
Galeano A, Santos S, Jimenez R, et~al (2024) Constraining non-negative matrix
  factorization to improve signature learning.
  \urlprefix\url{https://openreview.net/forum?id=AcGUW5655J}, openReview
  submission

\bibitem[{Galeano and Paccanaro(2022)}]{galeano2022}
Galeano D, Paccanaro A (2022) Machine learning prediction of side effects for
  drugs in clinical trials. Cell reports methods 2(12)

\bibitem[{Gillis(2015)}]{gillis2014}
Gillis N (2015) The why and how of nonnegative matrix factorization. In:
  Suykens JAK, Signoretto M, Argyriou A (eds) Regularization, Optimization,
  Kernels, and Support Vector Machines. Chapman \& Hall/CRC, Boca Raton, FL, p
  257--291

\bibitem[{Holzinger and Swineford(1939)}]{holzinger1939}
Holzinger KJ, Swineford F (1939) A Study in Factor Analysis: The Stability of a
  Bi-Factor Solution. No.~48 in Supplementary Educational Monographs,
  Department of Education, University of Chicago, Chicago

\bibitem[{Hwang and Takane(2004)}]{hwang2004}
Hwang H, Takane Y (2004) Generalized structured component analysis.
  Psychometrika 69(1):81--99. \doi{10.1007/BF02295841}

\bibitem[{J{\"o}reskog(1969)}]{joreskog1969}
J{\"o}reskog KG (1969) A general approach to confirmatory maximum likelihood
  factor analysis. Psychometrika 34(2):183--202. \doi{10.1007/BF02289343}

\bibitem[{Laurberg et~al(2008)Laurberg, Christensen, Plumbley, Hansen, and
  Jensen}]{Laurberg2008}
Laurberg H, Christensen MG, Plumbley MD, et~al (2008) Theorems on positive
  data: On the uniqueness of {NMF}. Computational Intelligence and Neuroscience
  2008:Article ID 764206. \doi{10.1155/2008/764206}, pMID: 18497868

\bibitem[{Lee and Seung(2000)}]{lee2000}
Lee D, Seung HS (2000) Algorithms for non-negative matrix factorization. In:
  Leen T, Dietterich T, Tresp V (eds) Advances in Neural Information Processing
  Systems 13, vol~13. MIT Press, pp 556--562,
  \urlprefix\url{https://proceedings.neurips.cc/paper_files/paper/2000/file/f9d1152547c0bde01830b7e8bd60024c-Paper.pdf}

\bibitem[{Lee and Seung(1999)}]{lee1999}
Lee DD, Seung HS (1999) Learning the parts of objects by non-negative matrix
  factorization. Nature 401(6755):788--791

\bibitem[{Leontief(1936)}]{leontief1936}
Leontief WW (1936) Quantitative input and output relations in the economic
  systems of the united states. The Review of Economic Statistics
  18(3):105--125

\bibitem[{Lodhia et~al(2023)Lodhia, H{\"u}tter, Uhler, and
  Zwiernik}]{lodhia2023}
Lodhia A, H{\"u}tter JC, Uhler C, et~al (2023) Positivity in linear gaussian
  structural equation models. arXiv preprint arXiv:230519884
  \urlprefix\url{https://arxiv.org/abs/2305.19884}

\bibitem[{Neale et~al(2016)Neale, Hunter, Pritikin, Zahery, Brick, Kirkpatrick,
  Estabrook, Bates, Maes, and Boker}]{openmx}
Neale MC, Hunter MD, Pritikin JN, et~al (2016) {OpenMx} 2.0: Extended
  structural equation and statistical modeling. Psychometrika 81(2):535--549.
  \doi{10.1007/s11336-014-9435-8}

\bibitem[{Paxton et~al(2011)Paxton, Hipp, and Marquart-Pyatt}]{paxton2011}
Paxton P, Hipp JR, Marquart-Pyatt S (2011) Nonrecursive Models: Endogeneity,
  Reciprocal Relationships, and Feedback Loops. SAGE, Thousand Oaks, CA

\bibitem[{Rosseel(2012)}]{lavaan}
Rosseel Y (2012) lavaan: An r package for structural equation modeling. Journal
  of statistical software 48(1):1--36

\bibitem[{Satoh(2025)}]{satoh2025lab}
Satoh K (2025) Applying non-negative matrix factorization with covariates to
  label matrix for classification.
  \urlprefix\url{https://arxiv.org/abs/2510.10375}, \eprint{2510.10375}

\bibitem[{Satoh(2026{\natexlab{a}})}]{satoh2025var}
Satoh K (2026{\natexlab{a}}) Applying non-negative matrix factorization with
  covariates to multivariate time series data as a vector autoregression model.
  Japanese Journal of Statistics and Data Science 9:79--97.
  \doi{10.1007/s42081-025-00314-0},
  \urlprefix\url{https://doi.org/10.1007/s42081-025-00314-0}

\bibitem[{Satoh(2026{\natexlab{b}})}]{satoh2025nmfkc}
Satoh K (2026{\natexlab{b}}) {nmfkc}: Non-Negative Matrix Factorization with
  Kernel Covariates. \doi{10.32614/CRAN.package.nmfkc},
  \urlprefix\url{https://CRAN.R-project.org/package=nmfkc}, r package version
  0.6.7

\bibitem[{Schmidt et~al(2009)Schmidt, Winther, and Hansen}]{schmidt2009}
Schmidt MN, Winther O, Hansen LK (2009) Bayesian non-negative matrix
  factorization. In: Independent Component Analysis and Signal Separation (ICA
  2009), Lecture Notes in Computer Science, vol 5441. Springer, Berlin,
  Heidelberg, pp 540--547, \doi{10.1007/978-3-642-00599-2_68}

\bibitem[{Stoffer and Poison(2026)}]{astsa}
Stoffer D, Poison N (2026) astsa: Applied Statistical Time Series Analysis.
  \urlprefix\url{https://CRAN.R-project.org/package=astsa}, r package version
  2.5; includes the \texttt{lap} dataset

\bibitem[{Tan and F{\'e}votte(2013)}]{TanFevotte2013}
Tan VYF, F{\'e}votte C (2013) Automatic relevance determination in nonnegative
  matrix factorization with the {$\beta$}-divergence. IEEE Transactions on
  Pattern Analysis and Machine Intelligence 35(7):1592--1605.
  \doi{10.1109/TPAMI.2012.240}, pMID: 23681989; arXiv:1111.6085

\bibitem[{Tenenhaus et~al(2005)Tenenhaus, Vinzi, Chatelin, and
  Lauro}]{tenenhaus2005}
Tenenhaus M, Vinzi VE, Chatelin YM, et~al (2005) {PLS} path modeling.
  Computational Statistics \& Data Analysis 48(1):159--205.
  \doi{10.1016/j.csda.2004.03.005}

\bibitem[{{University of Wisconsin Population Health
  Institute}(2025)}]{CHR2025}
{University of Wisconsin Population Health Institute} (2025) {County Health
  Rankings \& Roadmaps 2025 Analytic Dataset (Supplemental Data Release)}.
  \urlprefix\url{https://www.countyhealthrankings.org/health-data/methodology-and-sources/data-documentation},
  dataset accessed via https://www.countyhealthrankings.org. Includes
  Supplemental Data Release dated November 4, 2025

\bibitem[{Yamashita(2024)}]{yamashita2024}
Yamashita N (2024) Matrix decomposition approach for structural equation
  modeling as an alternative to covariance structure analysis and its
  theoretical properties. Structural Equation Modeling: A Multidisciplinary
  Journal 31(5):817--834. \doi{10.1080/10705511.2024.2342381}

\bibitem[{Zhang et~al(2016)Zhang, Tian, and Tang}]{zhang2016}
Zhang YQ, Tian GL, Tang NS (2016) Latent variable selection in structural
  equation models. Journal of Multivariate Analysis 152:190--205

\bibitem[{Zhou et~al(2022)Zhou, Ying, and Palomar}]{zhou2022}
Zhou R, Ying J, Palomar DP (2022) Covariance matrix estimation under low-rank
  factor model with nonnegative correlations. IEEE Transactions on Signal
  Processing 70:4020--4030

\end{thebibliography}

\end{document}